\documentclass[11pt]{article}
\usepackage{jcappub}
\usepackage{bm}
\usepackage{color}
\usepackage{graphicx}
\usepackage{multirow}
\usepackage{epstopdf}
\usepackage{geometry}

\newcommand{\qm}{\left (q \over m_N \right )}
\renewcommand{\qm}{\q}
\newcommand{\qmsq}{\qmsq}
\renewcommand{\qmsq}{\q^2}

\newcommand{\sdfrac}[2]{\mbox{\small$\displaystyle\frac{#1}{#2}$}}

\newcommand{\scdot}{\mkern-2mu\cdot\mkern-2mu}

\newcommand{\be}{\begin{equation}}
\newcommand{\ee}{\end{equation}}
\newcommand{\een}{\end{subequations}}
\newcommand{\ben}{\begin{subequations}}
\newcommand{\beq}{\begin{eqalignno}}
\newcommand{\eeq}{\end{eqalignno}}

\newcommand{\lsim}{\mathrel{\mathop{\kern 0pt \rlap
      {\raise.2ex\hbox{$<$}}}\lower.9ex\hbox{\kern-.190em $ \sim$}}}
\newcommand{\gsim}{\mathrel{\mathop{\kern 0pt
      \rlap{\raise.2ex\hbox{$>$}}}\lower.9ex\hbox{\kern-.190em $\sim$}}}

\newcommand{\VectorTypefaceArrow}{
\let\oldvec\vec
\renewcommand{\vec}[1]{\oldvec{##1}} 
\newcommand{\uvec}[1]{\hat{##1}} 
}

\VectorTypefaceArrow

\newcommand{\op}[1]{\widehat{#1}} 

\newcommand{\curr}[1]{#1} 

\usepackage{upgreek} 
\newcommand{\q}{{\widetilde{q}}}
\newcommand{\calO}{{\cal O}}

\usepackage{mathtools}
\newcommand{\myoverbracket}[1]{\overbracket[0.5pt][3pt]{#1}{}}

\newcommand{\plus}{\,+}

\usepackage{booktabs}
\usepackage{xcolor}
\usepackage{arydshln}

\allowdisplaybreaks

\title{The phenomenology of nuclear scattering for a WIMP of arbitrary spin}

\author[a]{Paolo Gondolo,}
\affiliation[a]{Department of Physics, University of Utah, 115 South 1400
  East \#201, Salt Lake City, Utah 84112-0830}
\emailAdd{paolo.gondolo@utah.edu}

\author[b]{Injun Jeong,}
\author[b]{Sunghyun Kang,}
\author[b]{Stefano Scopel,}
\author[c]{Gaurav Tomar}
\affiliation[b]{Department of Physics, Sogang University, Seoul 121-742, South Korea}
\affiliation[c]{Physik-Department, Technische Universit\"at M\"unchen, James-Franck-Stra\ss e, 85748
  Garching, Germany}
\emailAdd{natson@naver.com}
\emailAdd{francis735@naver.com}
\emailAdd{scopel@sogang.ac.kr}
\emailAdd{physics.tomar@tum.de}

\abstract{We provide a first systematic and quantitative discussion of
  the phenomenology of the non--relativistic effective Hamiltonian
  describing the nuclear scattering process for a Weakly Interacting
  Massive Particle (WIMP) of arbitrary spin $j_\chi$.  To this aim we
  obtain constraints from a representative sample of present direct
  detection experiments assuming the WIMP--nucleus scattering process
  to be driven by each one of the 44 effective couplings that arise
  for $j_\chi\le$ 2.  We find that a high value of the multipolarity
  $s\le 2 j_\chi$ of the coupling, related to the power of the
  momentum transfer $q$ appearing in the scattering amplitude, leads
  to a suppression of the expected rates and pushes the expected
  differential spectra to large recoil energies $E_R$. For $s\le$ 4
  the effective scales probed by direct detection experiments can be
  suppressed by up to 5 orders of magnitude compared to the case of a
  standard spin--independent interaction. For operators with large $s$
  the expected differential spectra can be pushed to recoil energies
  in the MeV range, with the largest part of the signal concentrated
  at $E_R\gsim$ 100 keV and a peculiar structure of peaks and minima
  arising when both the nuclear target and the WIMP are heavy. As a
  consequence the present bounds on the effective operators can be
  significantly improved by extending the recoil energy intervals to
  higher recoil energies.  Our analysis assumes effective interaction
  operators that are irreducible under the rotation group. Such
  operators drive the interactions of high--multipole dark matter
  candidates, i.e. states that possess only the highest multipole
  allowed by their spin. As a consequence our analysis represents also
  the first phenomenological study of the direct detection of
  quadrupolar, octupolar, and hexadecapolar dark matter.}

\begin{document}
\hspace*{107.5mm}{CQUEST-2021-0656}\\
\hspace*{116mm}{TUM-HEP 1311/20}
\maketitle
\section{Introduction}
\label{sec:introduction}

Weakly Interacting Massive Particles (WIMPs) are the most popular
particle candidates to provide the invisible halos of Galaxies,
including the Milky Way. WIMPs are expected to have only weak--type
interactions and to have a mass $m_\chi$ falling in the GeV--TeV
range. Direct detection (DD) experiments look for the interaction of
WIMPs with ordinary matter through the measurement of the recoil
energy $E_R\lsim$ 10 keV -- 100 keV imparted by WIMPs when they
scatter elastically off the nuclear targets of a wide range of
solid--state and liquid low--background underground detectors.

Model-independent approaches have become increasingly popular to
interpret Dark Matter (DM) search
experiments~\cite{chang_momentum_dependence_2010, dobrescu_nreft,
  fan_2010, Hisano_vector_effective, Hisano_EW_DM, hill_solon_nreft,
  peter_nreft, cirelli_tools_2013, effective_wimps_2014, catena_nreft,
  catena_directionality_nreft_2015, Hisano_Wino_DM,
  Catena_Gondolo_global_fits, cerdeno_nreft,
  Catena_Gondolo_global_limits, nreft_bayesian, xenon100_nreft,
  cresst_nreft,matching_solon1,matching_solon2,chiral_eft,hoferichter_si,bishara_2017,sogang_scaling_law_nr,
  dama_inelastic_eft_sogang} due to the growing tension between the
WIMPs arising in popular extensions of the Standard Model (SM) such as
Supersymmetry or Large Extra Dimensions and the constraints from the
Large Hadron Collider (LHC).  In particular, since the DD process is
non--relativistic (NR) the WIMP-nucleon interaction can be
parameterized with an effective Hamiltonian ${\bf\mathcal{H}}$ that
complies with Galilean symmetry.  The effective Hamiltonian
${\bf\mathcal{H}}$ to zero-th order in the WIMP-nucleon relative
velocity $\vec{v}$ and momentum transfer $\vec{q}$ has been known
since at least Ref.~\cite{GoodmanWitten}, and consists of the usual
spin-dependent (SD) and spin-independent (SI) terms.  To first order
in $\vec{v}$, the effective Hamiltonian ${\bf\mathcal{H}}$ has been
systematically described in~\cite{haxton1,haxton2} for WIMPs of spin 0
and 1/2, and less systematically described
in~\cite{krauss_spin_1,catena_krauss_spin_1} for WIMPs of spin 1 and
in~\cite{barger_2008} for WIMPs of spin 3/2.

Recently in\cite{all_spins} the NR effective Hamiltonian for
WIMP--nucleous scattering has been extended to include WIMPs of
arbitrary spin $j_\chi$ in the approximation of one--nucleon currents.
In Ref.~\cite{all_spins} ${\bf\mathcal{H}}$ is written in a complete
basis of rotationally invariant operators organized according to the
rank of the $2 j_\chi+1$ irreducible operator products of up to $2
j_\chi$ WIMP spin vectors.  In particular, for a WIMP of spin $j_\chi$
a basis of $4+20 j_\chi$ independent operators arise that can be
matched to any high-energy model of particle dark matter, including
elementary particles and composite states.

So far, only 20 of such operators for a WIMP with $j_\chi\le$1 have
been considered in the
literature~\cite{haxton1,haxton2,krauss_spin_1,catena_krauss_spin_1}. As
a consequence, the introduction in ~\cite{all_spins} of so many new
interaction terms for $j_\chi\ge$1 has potentially interesting
phenomenological consequences.

In particular, for each new interaction term a different scaling law
for the WIMP cross section, needed to compare the results of
experiments using different target nuclei, is expected to
arise. Moreover, the new high--rank operators that arise at high
values of the spin $j_\chi$ depend on increasing powers of the
transferred momentum $q\equiv |\vec{q}|$. On the one hand such
increasing momentum suppression implies lower expected rates, so that
present experimental sensitivities are expected to probe increasingly
low values of the effective energy scale of the effective
theory. Moreover, for a given choice of the WIMP velocity distribution
and $m_\chi\gsim$ 100 GeV the increasing powers of $q$ shift the
expected rate spectra to high recoil energies $E_R\simeq$1 MeV, much
higher than those usually expected.

The goal of the present paper is to provide a first systematic and
quantitative discussion of the phenomenological aspects outlined above
for the effective interaction operators introduced in~\cite{all_spins}
. To this aim, we will consider each of the 44 operators that can
arise in the nuclear scattering of a WIMP with $j_\chi\le$2.  Due to
the large dimensionality of the parameter space of high--spin DM,
starting with analyzing the phenomenology of one operator of such basis
at a time appears like a sensible approach.  In
Section~\ref{sec:high_multipole} we elaborate on the possibility that
such basis may have indeed a connection to the physical world, and that
DM candidates whose phenomenology is driven by such operators do
exist.

Throughout the paper we will assume a standard
Maxwellian in the Galactic rest frame cut at the WIMP escape velocity
$u_{esc}$=550 km/s and with reference values for the other relevant
parameters: $\rho_\chi$=0.3 GeV/cm$^3$ for the number density of the
WIMPs in the neighborhood of the Sun and $v_{rms}$=270 km/s for the
WIMP root--mean-square velocity.

The paper is organized as follows. In Section~\ref{sec:motivation} we
discuss the theoretical motivation of DM candidates of arbitrary spin
(Section~\ref{sec:high_spin}) and high multipole
(Section~\ref{sec:high_multipole}). In Section \ref{sec:eft} we
outline the results of Ref.~\cite{all_spins}, where the
non--relativistic effective operators for $j_\chi\ge$1 are introduced
and the analytic expressions of the relevant expected rates are
calculated. Section \ref{sec:sensitivities} is devoted to our
phenomenological discussion: in Section~\ref{sec:rate} we summarize
the expressions for the scattering rate; Section~\ref{sec:present}
provides a discussion of the values of the energy scale of the
effective theory probed by a representative selection of present
direct--detection experiments; in Section~\ref{sec:diff_rate} we
discuss the expected differential rate, showing that when the
scattering process is driven by a high--rank (high--spin) operator and
both the target and the WIMP are heavy the largest part of the signal
is concentrated at $E_R\gsim$ 100 keV with a characteristic pattern of
peaks and minima at high recoil energies; in
Section~\ref{sec:limit_improvements} we show how the present bounds
can be significantly improved by extending the recoil energy intervals
to higher recoil energies. Finally, we provide our Conclusions in
Section~\ref{sec:conclusions}. The details of the procedure followed
to obtain the upper bounds is described in the Appendix.

\section{High-spin and multipolar dark matter}
\label{sec:motivation}

In the present Section we discuss the theoretical motivation for DM
candidates of arbitrarily high spin. Moreover, we wish to elaborate on
the possibility that it possesses only the highest anomalous moment
allowed for its spin, implying that its interaction with
ordinary matter is driven by the high--rank irreducible operators
introduced in Eq.~(\ref{eq:basis_WIMP_nucleon_operators_alt}) and
whose phenomenology is the subject of our analysis.

\subsection{High--spin DM}
\label{sec:high_spin}

While most (but not all) of the theoretical and experimental work on
detection of particle dark matter has been focused on dark matter
particles that are elementary and have spin 0 or 1/2, there is no
compelling reason for dark matter particles to be elementary, or for
their spin to be limited to 0 and 1/2. In fact, as is well-known,
non-elementary particles and particles of spin higher than 1/2 exist
in nature. We include here not only the states of particle physics,
such as protons, neutrons, hadrons, gauge bosons, etc., but also
larger composite objects like atomic nuclei, atoms, molecules, etc.
In particular, the possibility to distinguish between pointlike and
non--pointlike dark-matter candidates in direct detection searches
through the shape of the nuclear recoil energy spectrum has been
discussed in the literature~\cite{kusenko_2002}.

Sometimes one hears that particles with high spin cannot have gauge
interactions, but this is an incorrect generalization of two quite
specific statements in the construction of interacting theories: there
are difficulties in coupling a massless~\cite{weinberg_high_spin} or
massive particle~\cite{velo_high_spin} of spin higher than 2 to
gravity and there are difficulties in coupling a massless particle to
the photon~\cite{weinberg_high_spin}. On the other hand there are no
difficulties in coupling any particle to photons at the effective
lagrangian level, i.e. in a theory valid at energy scales smaller than
the binding energy of the particle. Witness to this is the coupling of
complex atoms and molecules, and even macroscopic objects, to the
electromagnetic field. Even theoretically, consistent effective
Lagrangians coupling towers of particles of higher and higher spin to
gravity have been constructed as low energy limits of string theory
(see e.g.~\cite{argyres_nappi}). And consistent effective couplings of
particles to the electromagnetic field have been obtained as
four-dimensional reductions of field theories in spacetimes with extra
dimensions (see, e.g,~\cite{reduction_aragone, reduction_rindani1, reduction_rindani2}).
The non-local character of the fundamental
theories involved bypasses the limits of local theories with a single
particle of high spin.

Composite dark matter has been introduced for example in technicolor
theories (where a new strongly interacting sector is postulated) \cite{Weinberg1979,Susskind1979} or as
another example in mirror dark matter models (where dark copies of the
standard model particles are assumed, complete with dark protons, dark
neutrons, dark atomic nuclei, and so on; see for example \cite{Kolb:1985aa} and the review~\cite{mirror_dm_2014}). The continuing interest in composite dark matter can be gathered for example from the references in \cite{Dondi:2019olm}. In
most of these scenarios, the dark matter particles have spin 1/2 and
interact with the standard model through kinetic mixing of the dark
and visible photon, or the dark and visible Higgs bosons, or the dark
and visible neutrinos. Models of this kind have also been considered
in the so-called asymmetric dark matter
scenarios~\cite{Petraki_Volkas_2013}. Most but not
all mirror dark matter candidates are elementary particles of spin 0
or 1/2. An exception we found in the literature is the pangenesis
model of~\cite{pangenesis_2012} in which dark matter is atomic, being
a mirror hydrogen atom composed of a mirror proton and a mirror
electron.

There are stable light atomic nuclei with spin 3/2 ($^7$Li, $^{11}$Be,
$^{11}$B) and spin 3 ($^{10}$B), and heavier stable nuclei can have
spins up to 9/2 (e.g., $^{73}$Ge or $^{87}$Sr). While it may not be
easy to find a theoretical mechanism in which heavy dark nuclei would
be more abundant that light ones, so little is known about the
dynamics in the dark sector that it may well be possible that only
dark nuclei with high spin (1, 3/2, 2, ...) interact with ordinary
nuclei and are in principle detectable in direct dark matter search
experiments. For example, dark heavy elements such as O, Ne, N, C, and
Fe (besides H and He) have been considered in the context of direct
dark matter detection~\cite{mirror_dm_2004} with a long range
interaction mediated by a kinetically-mixed photon--dark photon
coupled to the electric charge of the dark nuclei.

Among the particles of spin 1, it is worth mentioning the deuteron and
the massive gauge bosons W and Z. The deuteron, which is stable, was
produced in the early universe during primordial nucleosynthesis. Its
primordial abundance was delicately determined by the particular
values of binding energies, lifetimes, and reaction cross sections of
hydrogen, deuterium, helium, and the neutron. An analog of the
deuteron in the dark sector may have a completely different, and
perhaps much larger abundance, thanks to a more favorable dynamics in
the dark sector; in fact, in models with mirror dark matter, the dark
primordial nucleosynthesis is set to happen at a slightly lower
temperature than in the SM sector, to avoid issues with the number of
neutrino species and other measures of the abundance of relativistic
particles in the early universe, and the lower nucleosynthesis
temperature produces a different pattern of primordial abundances,
with $^4$He being dominant over the other
species~\cite{dark_bbn_Berezhiani_2000}.

One can also imagine a ``quasi-mirror'' sector that is almost a copy
of the SM model but has small differences in the values of some
parameters (models with broken mirror symmetry can be found for
example in~\cite{Foot_Lew_LR_symmetric_1994,
  mirror_neutrinos_Berezhiani_1995, mirror_machos_Mohapatra_1999,
  Foot_two_vacua_2000}.  In a quasi-mirror dark sector the dark
neutron may be lighter than the dark proton. Indeed, the experimental
fact that the mass of the proton is slightly smaller than the mass of
the neutron is not well understood theoretically, as the
proton--neutron mass difference is moved one level lower into a mass
difference between up and down quarks, which is obtained
phenomenologically by fits to experimental data and for which there is
no theoretical calculation from first principles. Thus it may well be
that a slightly different dynamics in the dark sector produces a dark
neutron lighter than the dark proton. Then a free dark proton may
decay into a dark neutron, a dark positron, and a dark neutrino, but a
dark neutron bound in a dark nucleus may not decay, and the stability
of atomic nuclei would be completely changed. This would open the
possibility that dark neutrons form dark multineutron states. There is
in principle no obstacle for these dark multineutrons to have a large
value of their spin. In addition, given enough symmetry, many of the
lower electric and magnetic multipoles of the dark multineutron may
vanish, making the dominant dark multineutron--photon interaction a
multipole of high order in the dark photon--photon kinetic mixing
scenario.

\subsection{High-multipole DM}
\label{sec:high_multipole}

In Section~\ref{sec:sensitivities} we discuss the phenomenology of
high--spin DM states assuming that their interaction with ordinary
matter is driven by one of the irreducible operators introduced
in~\cite{all_spins} and given in
Eq.~(\ref{eq:basis_WIMP_nucleon_operators_alt}). In particular, for a
WIMP of spin $j_\chi$ in Section~\ref{sec:sensitivities} the
phenomenology of the operators of maximal rank $s$ = 2 $j_\chi$ is
shown (although such results can be easily rescaled to lower $s$
values using the coefficient plotted in Fig.~\ref{fig:b_chi_s}). As
discussed in Ref.~\cite{all_spins} the use of the operators of
Eq.~(\ref{eq:basis_WIMP_nucleon_operators_alt}) as a basis for the
effective Hamiltonian of the WIMP--nucleus interaction has several
advantages: for instance, it is complete, it avoids double counting
and simplifies the calculation of the cross section, that results in a
sum of cleanly separated contributions from irreducible operators of
different ranks that do not interfere. From this point of view,
however, to use such specific basis of operators instead of any other
appears to be a mere technicality, and one may argue that the
non--relativistic limit of a generic Hamiltonian should naturally
contain a mix of operators of all ranks, with the contribution of the
highest rank operators unavoidably suppressed. In this Section we wish
to elaborate on the possibility that, instead, such a basis may have
indeed a connection to the physical world, i.e. it is possible to
conceive DM candidates whose interaction with ordinary matter is
driven by the highest multipole moments connected to the high rank
operators whose phenomenology is the subject of our analysis.

There is a general relation between the spin of the particle and the
highest nonzero multipole moment it can possess. The Wigner-Eckart
theorem and analogous group theoretic arguments in special relativity,
allow a particle of spin--$j_\chi$ to have moments up to order $2
j_\chi+1$: a monopole for spin-0, a dipole for spin-1/2, etc. As a
real world example, the neutron has zero net electric charge, nonzero
magnetic moment, and all of its higher moments vanish.

Molecules in the dark sector provide an example of both compositeness
and high multipole moments. Dipolar molecules (having zero net charge
and non-zero permanent electric dipole moment) are well known, while
quadrupolar, octupolar, and hexadecapolar molecules may be less
known. Which multiple moment a molecule possesses is determined by the
kind of symmetry its distribution of charges has. Examples of
quadrupolar molecules, which have zero charge and zero dipole moment
but non-zero quadrupole moment, are the linear molecules of carbon
dioxide CO$_2$, CS$_2$ and the planar molecules of
C$_6$H$_6$. Among the octupolar molecules, whose first nonzero
multipole is the octupole, are molecules with tetrahedral symmetry
like CH$_4$ and CF$_4$. Hexadecapole molecules, which possess a
non-zero hexadecapole but have neither a dipole, quadrupole nor
octupole moment, include the octahedral symmetry molecules of SF$_6$
and UF$_6$.

It is possible to envisage a dynamics in the dark sector that may lead
to dark molecules of similarly high symmetry. It is important to
notice that for what concern the detection of dark matter molecules,
it is the charge distribution of the charges coupled to ordinary
standard model particles that matters, while the formation and
stability of dark molecules may be governed by forces that act within
the dark sector exclusively. In addition, one needs to consider the
phenomenon of induced polarization. In the case of multipolar dark
molecules in the vicinity of ordinary matter one may worry that
induced dipole (or other low) moments would come to dominate the
interaction at low energies. This may not happen either because of a
fortuitous arrangement of charges giving a small polarizability, or
more generally because the interaction range of the force between dark
and ordinary matter may be so small that there is no time for induced
multipoles to develop during the scattering of dark matter molecules
with ordinary matter.

The idea that dark matter may carry only multipoles of high order
generates the possibility that dark matter may be a neutral particle
that possesses only the highest anomalous moment allowed for its spin.
The highest multiple moment for a particle of spin $j_\chi$ is the
$2^{2 j_\chi}$ multipole, e.g., the dipole for spin-1/2, the
quadrupole for spin-1, the octupole for spin-3/2, and the hexadecapole
for spin-2. The anomalous moment could be an electric or magnetic
moment (examples from the literature are recalled later in this
section) or it could be a multipole moment in the interaction of dark
matter with new gauge fields in the dark sector, Abelian or
non-Abelian. One can thus have ``magnetic-dipole dark matter'' or
``electric-octupole dark matter'' or even higher multipole dark matter
according to which multipole moment is the lowest non-vanishing
moment.

For electromagnetic interactions and multipoles, effective Lagrangians
involve the electromagnetic field strength $F^{\mu\nu}$ multiplied by
an increasing number of derivatives of the particle fields, or of the
momentum exchange in the corresponding amplitude, multiplied by an
appropriate power of the particle radius or of the inverse of the
particle mass. It is important to stress that it is the particle mass,
or the particle radius, that accompany the nonrenormalizable operators
in a multipolar effective Lagrangian, and not an energy cutoff scale
at which the interactions become weak or strong, or at which new
degrees of freedom become dynamical. This is exemplified by the Fermi
weak theory on one hand and hadrons on the other, for which in the
first case the mediator mass is (much) larger than the energy scale
involved in the weak processes, and in the second case the scale of
strong interactions is lower than the mass of the hadrons. 

In absence of CP violation the highest anomalous multipole alternates
between electric and magnetic as the spin of the particle increases: a
magnetic dipole for spin 1/2, an electric quadrupole for spin 1, a
magnetic octupole for spin 3/2, an electric hexadecapole for spin 2,
and so on. If $F^{\mu\nu}$ is replaced by its dual $\tilde{F}^{\mu\nu}
= \epsilon_{\mu\nu\lambda\rho} F^{\lambda\rho}/2$, the multipoles in
the series switch character: an electric dipole for spin 1/2, a
magnetic quadrupole for spin 1, an electric octupole for spin 3/2, a
magnetic hexadecapole for spin 2, and so on.

For spin-1/2, a particle can have an electric and a magnetic dipole,
the latter with a contribution from the charge (the Dirac magnetic
moment) to which an extra contribution may be added (the anomalous
magnetic moment). The classic example of a neutral spin-1/2 particle
with nonzero anomalous magnetic moment is the neutron.  A dark neutron
analog for dark matter was suggested already
in~\cite{barger_marfatia_dipole, dipolar_dm_Masso_2009}, where dipolar
dark matter was introduced. In that case, the dark matter was
considered to interact with the usual electromagnetic field and have
zero electric charge but non-zero anomalous magnetic moment, or
non-zero electric dipole moment in case parity-violating interactions
were allowed.  The interaction lagrangian of a magnetic-dipole dark
matter particle $\chi$, of spin 1/2 and zero electric charge, reads
\begin{align}
{\cal L}_{\rm int} = - \frac{i e \kappa}{2m_\chi^2} \, \overline{\chi}  \sigma_{\mu\nu}  \chi \, F^{\mu\nu} ,
\label{eq:L_magnetic_dipole}
\end{align}
where $F^{\mu\nu}$ is the electromagnetic field strength tensor,
$\sigma_{\mu\nu}$ is the usual combination of Dirac $\gamma$ matrices,
$e$ the elementary unit charge (not the particle charge, which is set
to zero), and $\kappa$ is the anomalous magnetic moment of the
particle. The contributions of the anomalous magnetic dipole term in
Eq.~(\ref{eq:L_magnetic_dipole}) can be distinguished in the
three-particle $\chi(p_1)$-$\chi(p_2)$-$A_\mu(q)$ vertex (momentum
$p_1$ incoming, momenta $p_2$ and $q=p_1-p_2$ outgoing) by being those
terms that contain the highest allowed power of the momentum $q^\mu$
carried by the photon
\begin{align}
\frac{i e \kappa}{m_\chi^2} \sigma_{\mu\nu} q^\nu .
\end{align}

The same scenario can be set-up for particles of higher spin by
keeping only their highest possible multipole. The highest-multipole
terms in the three-particle $\chi$-$\chi$-$\gamma$ vertex can be
distinguished as being the terms that carry the $2j_\chi$-th power
(the highest power) of the photon momentum $q^\mu$ (the completely
symmetric homogenous combination of $2 j_\chi$ vectors $q^\mu$).

For spin-1, the W boson has a measured magnetic dipole and electric
quadrupole moments, in agreement with the gauge symmetry of the
Standard Model~\cite{pdg2020}. Its magnetic dipole moment is usually
parametrized as $\mu_W = e(1+\kappa+\lambda)/2m_W$, and its electric
quadrupole moment as $Q_W = -e(\kappa-\lambda)/m_W^2$, where the
parameters $\kappa$ and $\lambda$ appear in the effective
Lagrangian~\cite{w_boson_moments_Hagiwara_1986}:
\begin{align}
\mathcal{L}_{WW\gamma} = - i e (1+\kappa) F_{\mu\nu} W^{+\mu} W^{-\nu} - i \frac{e\lambda}{m_W^2}  F^{\nu\lambda} W^{+}_{\lambda\mu} W^{-\mu}_{\ \ \nu} .
\end{align}
In the Standard Model, $\kappa=1$ and $\lambda=0$.  The matrix element
for the elastic scattering of spin-1 quadrupolar dark matter by
one-photon exchange can be borrowed from the analogous matrix element
for a deuteron~\cite{deuteron_Arnold_1981}:
\begin{align}
\langle p_2\lambda_2 | j_\mu | p_1\lambda_1\rangle = \sdfrac{G_3(q^2)}{2m_\chi^2} \, (p_1+p_2)^\mu \, \epsilon_1 \scdot q \, \epsilon_2^* \scdot q,
\end{align}
where $j_\mu$ is the electromagnetic current, $p_1^\mu,\epsilon_1^\mu$
and $p_2^\mu,\epsilon_2^\mu$ are the incoming and outgoing momenta and
polarization vectors (of polarization $\lambda_1$ and $\lambda_2$) of
the dark matter particle, $m_\chi$ is the DM particle mass, and $q^\mu
= p_1^\mu - p_2^\mu$ is the momentum transfer.  Perturbative
expressions are known explicitly: for a quadrupolar particle
$\chi_\mu$ the Feynman rule for the
$\chi_\alpha(p_1)$-$\chi_\beta(p_2)$-$A_\mu(q)$ vertex (see
e.g.~\cite{gaemers_gounaris_1978}; the momentum $p_1$ is incoming, the
momenta $p_2$ and $q=p_1-p_2$ are outgoing)
\begin{align}
\frac{i e \lambda}{m_\chi^2} \, (p_1+p_2)^\mu \Big(\sdfrac{1}{2} q^2 g_{\alpha\beta} - q_\alpha q_\beta\Big),
\end{align}
and as a lagrangian term~\cite{baur_zeppenfeld_1989}
\begin{align}
{\cal L}_{\rm int} = - \frac{ie\lambda}{m_\chi^2} \, \chi^\dagger_{\mu\lambda} \chi^{\lambda}_{\ \nu} \, F^{\mu\nu},
\end{align}
where $\chi_{\mu\nu} = \partial_\mu \chi_\nu - \partial_\nu \chi_\mu$. Here $\lambda$ is the anomalous quadrupole moment of the spin-1 particle.

In principle, the above considerations on anomalous multipole dark matter are not limited to the exchange of a massless vector mediator. A massive mediator would present the same structure for the interaction lagrangian and the interaction vertex, and would provide short range interactions. A scalar or fermion mediator, whether massive or massless, would also be subjected to a multipole expansion, as the multipole expansion is basically an expansion of a potential in spherical harmonics, or essentially an expansion in (symmetric traceless combinations of) derivatives of the mediator field. 

We are not aware of any study of multipolar dark matter beyond the dipole. This paper provides the first phenomenological study of the direct detection of quadrupolar, octupolar, and hexadecapolar dark matter.

\section{Effective theory of nuclear scattering for a WIMP of arbitrary spin}
\label{sec:eft}

In Ref.~\cite{all_spins} a systematic approach is introduced to
characterize the most general non--relativistic WIMP--nucleus
interaction allowed by Galilean invariance for a WIMP of arbitrary
spin $j_\chi$ in the approximation of one--nucleon currents,
i.e. assuming that the WIMP interacts with one nucleon at at time.
The procedure consists in organizing the WIMP currents according to
the rank of the 2$j_\chi$ + 1 irreducible operator products of the up
to the $2 j_\chi$ WIMP spin vectors $\vec{S}_\chi$ required to mediate
transitions where the third component of the WIMP spin changes from
$\pm j_\chi$ to $\mp j_\chi$. Using index notation $S_i$ for the
$i$-th component of the vector $\vec{S}_\chi$ (and dropping the
subscript $\chi$ in $S_{\chi,i}$ for more readability), these
interaction terms contain no $S_i$ or a product of $s$ factors $S_i$
up to $s=2j_\chi$,
\begin{align}
1, \quad S_{i_1}, \quad S_{i_1} S_{i_2}, \quad S_{i_1}S_{i_2}S_{i_3}, \quad\dots, \quad S_{i_1} S_{i_2} \cdots S_{i_{2j_\chi}}.
\label{eq:products_of_S}
\end{align}

The calculation of the cross section for WIMP--nucleus scattering is
greatly simplified if for the products of WIMP spin operators one uses
irreducible tensors (i.e., belonging to irreducible representations of
the rotation group). Irreducible tensors are completely symmetric
under exchange of any two of their indices and have zero trace under
contraction of any number of pairs of indices (they are symmetric
traceless tensors). In particular irreducible tensor operators of
different rank are independent, in the sense that the trace of their
product is zero, so that there are no interference terms in the cross
section between irreducible operators of different spin. Therefore,
the products of Eq.~(\ref{eq:products_of_S}) are conveniently
substituted by the following $2j_\chi+1$ irreducible spin tensors:

\begin{align}
  1, \quad \myoverbracket{S_{i_1}}, \quad \myoverbracket{S_{i_1} S_{i_2}}, \quad \myoverbracket{S_{i_1}S_{i_2}S_{i_3}}, \quad\dots, \quad \myoverbracket{S_{i_1} S_{i_2} \cdots S_{i_{2j_\chi}}},
  \label{eq:irreducible_products_of_S}
\end{align}

\noindent where we use an overbracket over an expression containing a
set of indices to indicate that the free indices under the bracket are
completely symmetrized and all of their contractions are subtracted.

Besides being a convenient basis to describe the WIMP current in spin
space the symmetric traceless products of spin operators of
Eq.~(\ref{eq:irreducible_products_of_S}) can also be seen as the
interaction terms appearing in the multipole expansion of the
WIMP-nucleus effective potential. For instance, given a scalar
potential $V(r_{\chi N})$ (with $r_{\chi N}$ the WIMP-nucleus
distance) the symmetric traceless combinations of $l$ of its derivatives
$\myoverbracket{\partial_{i_1} \partial_{i_2} \cdots \partial_{i_s}}
V$ represents the $l$-th multipole in its multipole expansion, and when
coupled to a generic combination of WIMP spin operators singles out
the one with the highest multipole:

\begin{align}
S_{i_1} S_{i_2} \cdots
S_{i_s} \,\myoverbracket{\partial_{i_1} \partial_{i_2} \cdots
  \partial_{i_s}} V(r_{\chi N})= \myoverbracket{ S_{i_1} S_{i_2}
  \cdots S_{i_s}} \,\myoverbracket{\partial_{i_1} \partial_{i_2}
  \cdots \partial_{i_s}} V(r_{\chi N})
\end{align}

Analogous expansions in derivatives of vector potentials
exist~\cite{all_spins}.  So operators that are irreducible tensors
under the rotation group can drive the effective interaction of the
high--multipole DM candidates discussed in
Section~\ref{sec:high_multipole}.

Once the number of WIMP spin factors is fixed to $s$, one couples the
WIMP currents to one of the five nucleon currents that arise in
WIMP--nucleon scattering from the nonrelativistic limit of the free
nucleon Dirac bilinears $\overline{\psi}_{\rm f} \Gamma \psi_{\rm i}$
(with $\Gamma$ any combination of Dirac $\gamma$ matrices and $\psi$
the Dirac spinor for a relativistic free nucleon) and the most general
interaction Hamiltonian depending on the WIMP spin operator and 
the nucleon spin operator:

\begin{align}
\op{\calO}_{M} = 1 ,
\quad 
\op{\vec{\calO}}_{\Sigma} = \vec{\sigma}_N ,
\quad
\op{\vec{\calO}}_{\Delta} = \vec{v}{}^{\plus}_{\chi N} ,
\quad
\op{\vec{\calO}}_{\Phi} = \vec{v}{}^{\plus}_{\chi N} \times \vec{\sigma}_N ,
\quad
\op{\calO}_{\Omega} =  \vec{v}{}^{\plus}_{\chi N}  \cdot \vec{\sigma}_N .
\label{eq:calO_X}
\end{align}
\noindent In the equation above $\vec{\sigma}_N$ is the vector of
Pauli spin matrices acting on the spin states of the nucleon $N$,
while

\begin{align}
  \vec{v}{}^{\plus}_{\chi N} = \vec{v}{}^{\plus}_{\chi} - \vec{v}{}^{\plus}_{N}
  \label{eq:v_perp_chi_N}
\end{align}

\noindent with:

\begin{align}
  \op{\vec{v}}{}^{\plus}_N = - \frac{i}{m_N} \left( \overrightarrow{\frac{\partial}{\partial\vec{r}_N}} - \overleftarrow{\frac{\partial}{\partial\vec{r}_N}}  \right) \label{eq:vperp}
  \end{align}
  \noindent (in the position representation), where $\vec{r}_N$ and
  $m_N$ are the position vector and the mass of the nucleon $N$, while
\begin{align}
\vec{v}{}^{\plus}_{\chi} = \vec{v}_{\chi} - \frac{ \vec{q}}{2m_\chi}. 
\label{eq:v_plus_chi}
\end{align}

\noindent For each operator $\op{O}_{X}$ ($X$=1, $\Sigma$, $\Delta$,
$\Phi$, $\Omega$) the number of $\q\equiv q/m_N$ factors is
constrained by rotational invariance. In particular, in the case of a
scalar nucleon operator $\op{O}_{X}$ ($X=M, {\Omega}$), the WIMP
operator $\op{o}$ must be a scalar, and all the indices $i_1\cdots
i_s$ in $S_{i_1} \cdots S_{i_s}$ must be saturated by terms
$\q_{i_1}\cdots \q_{i_s}$. The resulting WIMP operator is $S_{i_1}
\cdots S_{i_s}\q_{i_1}\cdots \q_{i_s}$. On the other hand, in the case
of a vector nucleon operator $\vec{\op{O}}_{X}$
($X=\Sigma,\Delta,\Phi$), a vector WIMP operator $\op{\vec{o}}$ is
needed, and the $s$ indices in $S_{i_1} \cdots S_{i_s}$ must be
saturated by an appropriate number of $\q$ factors in order to obtain
a vector. This can be achieved in three ways: (1) by using $s-1$
factors of $\q$ to produce $S_{i_1} \cdots S_{i_s}\q_{i_1}\cdots
\q_{i_{s-1}}$ with free index $i_s$, (2) by using $s$ factors of $\q$
to produce $\epsilon_{i_s l m} S_{i_1} \cdots S_{i_{s-1}} S_l
\q_{i_1}\cdots \q_{i_{s-1}}\q_m$, again with free index $i_s$, and (3)
by using $s+1$ factors of $\q$ to produce $S_{i_1} \cdots
S_{i_s}\q_{i_1}\cdots \q_{i_s}\q_{i_{s+1}}$, with free index
$i_{s+1}$.  Using the irreducible spin products in
Eq.~(\ref{eq:irreducible_products_of_S})
in place of those in Eq.~(\ref{eq:products_of_S}), one introduces the scalar WIMP operators
\begin{align}
i^s \myoverbracket{ S_{i_1} \cdots S_{i_s} } \, \q_{i_1} \cdots \q_{i_s} ,
\end{align}
and the vector WIMP operators
\begin{align}
i^s \myoverbracket{S_{i_1} \cdots S_{i_s}}\q_{i_1}\cdots \q_{i_{s-1}} && \text{(free index $i_s$)},
\nonumber \\
i^s \epsilon_{ijk}  \myoverbracket{S_{i_1} \cdots S_{i_{s-1}} S_{j}}
\q_{i_1}\cdots \q_{i_{s-1}}\q_{k} && \text{(free index $i$)},
 \nonumber\\
i^{s+1} \myoverbracket{ S_{i_1} \cdots S_{i_s}} \q_{i_1}\cdots \q_{i_s}\q_{i_{s+1}} && \text{(free index $i_{s+1}$)},
\end{align}

\noindent so that the following basis
of WIMP--nucleon operators $\calO_{X,s,l}$, all of which are
irreducible in WIMP spin space and Hermitian, is obtained:

\begin{align}
\calO_{M,s,s} & = \myoverbracket{ (i \, \vec{\q} \cdot \vec{S}_\chi)^s } && (s\ge0),\nonumber \\
\calO_{\Omega,s,s} & = \myoverbracket{ (i \, \vec{\q} \cdot \vec{S}_\chi)^s } \, (\vec{v}{}^{\plus}_{\chi N} \cdot \vec{S}_N) && (s\ge0),\nonumber \\
\calO_{\Sigma,s,s-1} & = \myoverbracket{ (i \,\vec{\q} \cdot \vec{S}_\chi)^{s-1} (\vec{S}_N \cdot \vec{S}_\chi) }  && (s\ge1), \nonumber\\
\calO_{\Sigma,s,s} & =\myoverbracket{ (i \,\vec{\q} \cdot \vec{S}_\chi)^{s-1} (i \, \vec{\q} \times \vec{S}_N \cdot \vec{S}_\chi) }  && (s\ge1),\nonumber \\
\calO_{\Sigma,s,s+1} & =\myoverbracket{ (i \, \vec{\q} \cdot \vec{S}_\chi)^s } \, (i \, \vec{\q} \cdot \vec{S}_N) && (s\ge0),\nonumber \\
\calO_{\Delta,s,s-1} & = \myoverbracket{ (i \,\vec{\q} \cdot \vec{S}_\chi)^{s-1} (\vec{v}{}^{\plus}_{\chi N} \cdot \vec{S}_\chi) }   && (s\ge1),\nonumber \\
\calO_{\Delta,s,s} & = \myoverbracket{ (i \,\vec{\q} \cdot \vec{S}_\chi)^{s-1} (i \, \vec{\q} \times  \vec{v}{}^{\plus}_{\chi N}  \cdot \vec{S}_\chi) }  && (s\ge1),\nonumber \\
\calO_{\Delta,s,s+1} & =\myoverbracket{ (i \, \vec{\q} \cdot \vec{S}_\chi)^s } (i\, \vec{\q} \cdot \vec{v}{}^{\plus}_{\chi N} )&& (s\ge0), \nonumber\\
\calO_{\Phi,s,s-1} & = \myoverbracket{ (i \,\vec{\q} \cdot \vec{S}_\chi)^{s-1} ( \vec{v}{}^{\plus}_{\chi N}  \times \vec{S}_N \cdot \vec{S}_\chi) }  && (s\ge1),\nonumber \\
\calO_{\Phi,s,s} & =  \myoverbracket{ (i \,\vec{\q} \cdot \vec{S}_\chi)^{s-1} (\vec{v}{}^{\plus}_{\chi N} \cdot \vec{S}_\chi) }  \, (i \, \vec{\q} \cdot \vec{S}_N) && (s\ge1), \nonumber\\
\calO_{\Phi,s,s+1} & = \myoverbracket{ (i \, \vec{\q} \cdot \vec{S}_\chi)^s } \, (i \, \vec{\q} \times \vec{v}{}^{\plus}_{\chi N} \cdot \vec{S}_N) && (s\ge0).
\label{eq:basis_WIMP_nucleon_operators_alt}
\end{align}

In terms of the Wilson coefficients $c^{\tau}_{X,s,l}(q)$
the WIMP--nucleon effective Hamiltonian ${\cal H}$ is a linear
combination of the WIMP--nucleon operators listed above:

\begin{align}
& {\cal H} = \sum_{X\tau s\,l} c^{\tau}_{X,s,l}(q) \, \calO_{X,s,l} \, t^\tau_N,
\label{eq:O_chi_N}
\end{align}

\noindent with $\tau$ an isospin index (0 for isoscalar and 1 for
isovector) and $t^0=1$, $t^1=\tau_3$ are nucleon isospin operators
(the $2\times2$ identity and the third Pauli matrix, respectively).
The relations between the isoscalar and isovector coupling constants
$c^0_{X,s,l}$ and $c^{1}_{X,s,l}$ and the proton and neutron coupling
  constants $c^{p}_{X,s,l}$ and $c^{n}_{X,s,l}$ is:
\begin{align}
c^{p}_{X,s,l}=\frac{c^{0}_{X,s,l}+c^{1}_{X,s,l}}{2} , \qquad
c^{n}_{X,s,l}=\frac{c^{0}_{X,s,l}-c^{1}_{X,s,l}}{2} .
\end{align}

In Table ~\ref{tab:Haxton_operators} we provide a dictionary between
the operator basis used in the literature for a WIMP of spin
$j_\chi\le$1~\cite{haxton1,haxton2,krauss_spin_1,catena_krauss_spin_1}
and the basis introduced in~\cite{all_spins} and summarized in
Eq.~(\ref{eq:basis_WIMP_nucleon_operators_alt}).

{ 
\begin{table}[t]\centering
  \caption{Non-relativistic Galilean invariant operators discussed in
    the literature (\cite{haxton2, krauss_spin_1,
      catena_krauss_spin_1}) for a dark matter particle of spin $0$,
    $1/2$ and $1$, and their relation with the WIMP--nucleon operators
    $\calO_{X,s,l}$ defined in
    Eqs.~(\ref{eq:basis_WIMP_nucleon_operators_alt}). Notice that the sign
    convention for the momentum transfer $\vec{q}$ used in this table
    and throughout the paper is opposite to that of
    Refs.~\cite{haxton2, krauss_spin_1, catena_krauss_spin_1}.}
\label{tab:Haxton_operators}
\renewcommand{\arraystretch}{1.5}
\addtolength{\tabcolsep}{2.0pt}
\vskip\baselineskip
\hspace{-4em}
\begin{minipage}{0.4\textwidth}
\begin{tabular}{@{}llr@{}}
\toprule
$\calO_{1}$ & $1$ & $\calO_{M,0,0}$ \\
[0.5ex]\cdashline{1-3}
$\calO_{2}$ & $(\vec{v}{}^{\plus}_{\chi N})^2 $ & $N.A.$ \\
[0.5ex]\cdashline{1-3}
$\calO_{3}$ & $-i \vec{S}_N \cdot ( \vec{\q} \times \vec{v}{}^{\plus}_{\chi N} ) $ & $-\calO_{\Phi,0,1}$ \\
[0.5ex]\cdashline{1-3}
$\calO_{4}$ & $\vec{S}_\chi \cdot \vec{S}_N$ & $\calO_{\Sigma,1,0}$ \\
[0.5ex]\cdashline{1-3}
$\calO_{5}$ & $ - i \vec{S}_\chi \cdot ( \vec{\q} \times \vec{v}{}^{\plus}_{\chi N} )$ & $-\calO_{\Delta,1,1}$ \\
[0.5ex]\cdashline{1-3}
$\calO_{6}$ & $(\vec{S}_\chi\cdot \vec{\q})  ( \vec{S}_N \cdot \vec{\q} )$ & $-\calO_{\Sigma,1,2}$ \\
[0.5ex]\cdashline{1-3}
$\calO_{7}$ & $ \vec{S}_N \cdot \vec{v}{}^{\plus}_{\chi N} $ & $\calO_{\Omega,0,0}$ \\
[0.5ex]\cdashline{1-3}
$\calO_{8}$ & $\vec{S}_\chi \cdot \vec{v}{}^{\plus}_{\chi N}$ & $\calO_{\Delta,1,0}$ \\
[0.5ex]\cdashline{1-3}
$\calO_{9}$ & $-i \vec{S}_\chi \cdot (\vec{S}_N \times \vec{\q} )$ & $\calO_{\Sigma,1,1}$ \\
[0.5ex]\cdashline{1-3}
$\calO_{10}$ & $-i \vec{S}_N \cdot \vec{\q}$ & $-\calO_{\Sigma,0,1}$ \\
[0.5ex]\cdashline{1-3}
$\calO_{11}$ & $-i \vec{S}_\chi \cdot \vec{\q}$ & $-\calO_{M,1,1}$ \\
[0.5ex]\cdashline{1-3}
$\calO_{12}$ & $\vec{S}_\chi \cdot (\vec{S}_N \times \vec{v}{}^{\plus}_{\chi N} )$ & $-\calO_{\Phi,1,0}$ \\
\bottomrule
\end{tabular}
\end{minipage}
\hspace{1em}
\begin{minipage}{0.4\textwidth}
\begin{tabular}{@{}llr@{}}
\toprule
$\calO_{13}$ & $\calO_{10}\calO_{8}$ & $-\calO_{\Phi,1,1}$ \\
[0.5ex]\cdashline{1-3}
$\calO_{14}$ & $\calO_{11}\calO_{7}$ & $-\calO_{\Omega,1,1}$ \\
[0.5ex]\cdashline{1-3}
$\calO_{15}$ & $-\calO_{11}\calO_{3}$ & $-\calO_{\Phi,1,2}$ \\
[0.5ex]\cdashline{1-3}
$\calO_{16}$ & $-\calO_{10}\calO_{5}$ &$-\calO_{\Phi,1,2}- \tilde{q}^2 \calO_{\Phi,1,0}$\\
[0.5ex]\cdashline{1-3}
$\calO_{17}$ & $-i\vec{\tilde{q}}\cdot{\bf \cal S}\cdot\vec{v}{}^{\plus}_{\chi N}$ & $\calO_{\Delta,2,1}$\\
[0.5ex]\cdashline{1-3}
$\calO_{18}$ & $-i\vec{\tilde{q}}\cdot{\bf \cal S}\cdot\vec{S}_N$ & $\calO_{\Sigma,2,1}-\frac{1}{3}\calO_{\Sigma,0,1}$\\
[0.5ex]\cdashline{1-3}
$\calO_{19}$ & $\vec{\tilde{q}}\cdot{\bf \cal S}\cdot\vec{\tilde{q}}$ & $\calO_{M,2,2}+\frac{1}{3}\tilde{q}^2\calO_{M,0,0}$\\
[0.5ex]\cdashline{1-3}
$\calO_{20}$ & $\left (\vec{S}_N \times \vec{\tilde{q}}\right )\cdot{\bf \cal S}\cdot\vec{\tilde{q}}$ & -$\calO_{\Sigma,2,2}$\\
[0.5ex]\cdashline{1-3}
$\calO_{21}$ & $\vec{v}{}^{\plus}_{\chi N}\cdot{\bf \cal S}\cdot\vec{S}_N$ & $\frac{1}{3}{\cal O}_{\Omega,0,0}$\\
[0.5ex]\cdashline{1-3}
$\calO_{22}$ & $\left (- i\vec{\tilde{q}}\times\vec{v}{}^{\plus}_{\chi N}\right )\cdot{\cal S}\cdot \vec{S}_N$ & $- {\cal O}_{\Phi,2,1}-\frac{1}{3}{\cal O}_{\Phi,0,1}$\\
[0.5ex]\cdashline{1-3}
$\calO_{23}$ & $- i\vec{\tilde{q}}\cdot{\cal S}\cdot\left (\vec{S}_N\times\vec{v}{}^{\plus}_{\chi N} \right )$ & $- {\cal O}_{\Phi,2,1}+\frac{1}{3}{\cal O}_{\Phi,0,1}$\\
[0.5ex]\cdashline{1-3}
$\calO_{24}$ & $- \vec{v}{}^{\plus}_{\chi N}\cdot{\cal S}\cdot\left (\vec{S}_N\times i\vec{\tilde{q}}\right)$ & $- {\cal O}_{\Phi,2,1}-\frac{1}{3}{\cal O}_{\Phi,0,1}$\\
\bottomrule
\end{tabular}
\vspace{1.5\baselineskip}
\end{minipage}
\end{table}
} 

The unpolarized differential cross section for WIMP-nucleus scattering
is given by the expression:

\be \frac{d\sigma_T}{d E_R}=\frac{2
  m_T}{2J_i+1}\frac{1}{v_{\chi T}^2}
\sum_{\tau=0,1}\sum_{\tau^{\prime}=0,1} \sum_{X}
R_X^{\tau\tau^{\prime}}\big(v^{\plus 2}_{\chi T}, \qmsq\big)
\sum_{J_f} \, \widetilde{W}^{\tau\tau^\prime}_{T\curr{X}} (q),
\label{eq:dsigma_de}
\ee

where $v_{\chi T}\equiv |\vec{v}_{\chi T}|$ is the WIMP speed in
the reference frame of the nucleus center of mass and:
\begin{align}
  \vec{v}{}^{\plus}_{\chi T} = \vec{v}{}^{\plus}_{\chi} - \vec{v}{}^{\plus}_{T}=\vec{v}_{\chi T}-\frac{\vec{q}}{2\mu_{\chi T}},
  \label{eq:v_chi_t}
\end{align}

\noindent with $\mu_{\chi T}$ the reduced WIMP--nucleus mass. The
sum in Eq.~(\ref{eq:dsigma_de}) is over $X=M$, $\Phi^{\prime \prime}$,
$\Phi^{\prime \prime} M$, $\tilde{\Phi}^\prime$, $\Sigma^{\prime
  \prime}$, $\Sigma^\prime$, $\Delta$, $\Delta\Sigma^\prime$ while:

\begin{align}
& \widetilde{W}_{TX}^{\tau\tau^{\prime}}(q) =
  W_{TX}^{\tau\tau^{\prime}}(q) , && \text{for $X=M,\Sigma',\Sigma''$}
  , \nonumber \\ & \widetilde{W}_{TX}^{\tau\tau^{\prime}}(q) = \q^2
  \, W_{TX}^{\tau\tau^{\prime}}(q) ,&& \text{for
    $X=\Delta,\widetilde\Phi',\Phi'',\Sigma'\Delta,\Phi''M$},
\label{eq:f_tilde}
\end{align}

\noindent and the nuclear response functions $W_{TX}$ are available in
the literature for the most common nuclear targets used in direct
detection experiments~\cite{haxton2, catena}. Notice that they do not
depend on $j_\chi$ thanks to the factorization between WIMP and
nuclear currents valid in one--nucleon approximation. On the other
hand the WIMP response functions $R_X^{\tau\tau^{\prime}}$ are given
by~\cite{all_spins}:

\begin{align}
R_{M}^{\tau \tau^\prime}&\big(v^{\plus 2}_{\chi T}, \qmsq\big) =
   v^{\plus 2}_{\chi T} \, R_{\Delta}^{\tau \tau^\prime}\big(v^{\plus 2}_{\chi T}, \qmsq\big) 
   + \sum_{s=0}^{2j_\chi} B_{j_\chi,s} c_{M,s,s}^\tau c_{M,s,s}^{\tau^\prime *}\qm^{2s} \nonumber
\\
R_{\Phi^{\prime \prime}}^{\tau \tau^\prime}&\big(v^{\plus 2}_{\chi T}, \qmsq\big) =
   \frac{1}{4}c_{\Phi,0,1}^{\tau} c_{\Phi,0,1}^{\tau^\prime *} \qm^{2} 
   \nonumber \\ &
   + \frac{1}{4}\sum_{s=1}^{2j_\chi} B_{j_\chi,s}  \qm^{2s-2} 
   \big ( c_{\Phi,s,s-1}^{\tau} - c_{\Phi,s,s+1}^{\tau} \qm^{2} \big )
   \big ( c_{\Phi,s,s-1}^{\tau^\prime *} - c_{\Phi,s,s+1}^{\tau^\prime *} \qm^{2} \big )
 \nonumber\\
 R_{\Phi^{\prime \prime} M}^{\tau \tau^\prime}&\big(v^{\plus 2}_{\chi T}, \qmsq\big) = 
   -  c_{\Phi,0,1}^{\tau} c_{M,0,0}^{\tau^\prime *} 
   + \sum_{s=1}^{2j_\chi} B_{j_\chi,s}  \qm^{2s-2}  
   \big ( c_{\Phi,s,s-1}^{\tau} - c_{\Phi,s,s+1}^{\tau} \qm^{2} \big ) c_{M,s,s}^{\tau^\prime *} ,
\nonumber\\
R_{\tilde{\Phi}^\prime}^{\tau \tau^\prime}&\big(v^{\plus 2}_{\chi T}, \qmsq\big) =
   \sum_{s=1}^{2j_\chi} B_{j_\chi,s} \frac{s+1}{8s}  \qm^{2s-2}\big (
   c_{\Phi,s,s-1}^{\tau} c_{\Phi,s,s-1}^{\tau^\prime *}
   + c_{\Phi,s,s}^{\tau} c_{\Phi,s,s}^{\tau^\prime *} \qm^{2}
  \big ) ,
\nonumber\\
R_{\Sigma^{\prime \prime}}^{\tau \tau^\prime}&\big(v^{\plus 2}_{\chi T}, \qmsq\big)  =
   v^{\plus 2}_{\chi T} \, R_{\tilde{\Phi}^\prime}^{\tau \tau^\prime}\left(v^{\plus 2}_{\chi T}, \qmsq\right) 
   + \frac{1}{4} c_{\Sigma,0,1}^{\tau} c_{\Sigma,0,1}^{\tau^\prime *} \qm^{2} 
   \nonumber \\ &
   + \sum_{s=1}^{2j_\chi} \frac{1}{4} B_{j_\chi,s}  \qm^{2s-2} 
   \big ( c_{\Sigma,s,s-1}^{\tau} - c_{\Sigma,s,s+1}^{\tau} \qm^{2} \big )
   \big ( c_{\Sigma,s,s-1}^{\tau^\prime *} - c_{\Sigma,s,s+1}^{\tau^\prime *} \qm^{2} \big )
   ,
\nonumber\\
R_{\Sigma^\prime}^{\tau \tau^\prime}&\big(v^{\plus 2}_{\chi T}, \qmsq\big)  =
   \frac{1}{2} \, v^{\plus 2}_{\chi T} \, R_{\Phi^{\prime \prime}}^{\tau \tau^\prime}\left(v^{\plus 2}_{\chi T}, \qmsq\right) 
   + \sum_{s=0}^{2j_\chi} \frac{1}{8} B_{j_\chi,s} \,
   c_{\Omega,s,s}^{\tau} c_{\Omega,s,s}^{\tau^\prime *}  v^{\plus 2}_{\chi T} \qm^{2s}
   \nonumber \\ & 
   + \sum_{s=1}^{2j_\chi}  \frac{1}{8} B_{j_\chi,s}  \frac{s+1}{s} \qm^{2s-2} 
   \big (c_{\Sigma,s,s-1}^{\tau} c_{\Sigma,s,s-1}^{\tau^\prime *} + c_{\Sigma,s,s}^{\tau} c_{\Sigma,s,s}^{\tau^\prime *} \qm^{2}  \big ) ,
\nonumber\\      
R_{\Delta}^{\tau \tau^\prime}&\big(v^{\plus 2}_{\chi T}, \qmsq\big) =
    \sum_{s=1}^{2j_\chi} 
    B_{j_\chi,s} \frac{s+1}{2s}  \qm^{2s-2} \big (
    c_{\Delta,s,s-1}^{\tau} c_{\Delta,s,s-1}^{\tau^\prime *}
    + c_{\Delta,s,s}^{\tau} c_{\Delta,s,s}^{\tau^\prime *} \qm^{2}
    \big ) ,
\nonumber \\
R_{\Delta \Sigma^\prime}^{\tau \tau^\prime}&\big(v^{\plus 2}_{\chi T}, \qmsq\big) =
    -\sum_{s=1}^{2j_\chi} B_{j_\chi,s} \frac{s+1}{2s} \qm^{2s-2} \big (
    c_{\Delta,s,s}^{\tau} c_{\Sigma,s,s-1}^{\tau^\prime *} 
    + c_{\Delta,s,s-1}^{\tau} c_{\Sigma,s,s}^{\tau^\prime *} 
    \big ),
    \label{eq:wimp_response_functions_sent}
\end{align}

\noindent where:

\begin{equation}
  B_{j_\chi,s}=\frac{s!}{(2s+1)!!}\frac{s!}{(2s-1)!!} \, K_{j_\chi,0} \cdots K_{j_\chi,s-1}
  \label{eq:b_jchi_s}
\end{equation}
with
\begin{equation}
  K_{j_\chi,i}=j_\chi\left(j_\chi+1\right)-\frac{i}{2}\left(\frac{i}{2}+1\right).
\end{equation}

\begin{figure}
\begin{center}
  \includegraphics[width=0.7\textwidth]{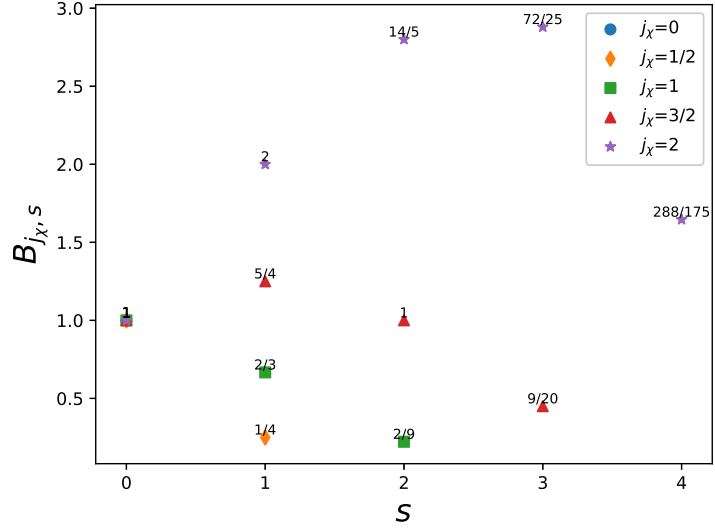}
\end{center}
\caption{Factor $B_{j_\chi,s}$ as given in Eq.~(\ref{eq:b_jchi_s}) as
  a function of $s$ and for $j_\chi$=0,1/2,1,2/3 and 2. For each
  operator $\calO_{X,s,l}$ of
  Eq.~(\ref{eq:basis_WIMP_nucleon_operators_alt}) this factor encodes the
  dependence on the WIMP spin $j_\chi$ of the response functions
  $R_X^{\tau\tau^{\prime}}$ in Eq.~(\ref{eq:wimp_response_functions_sent}).
\label{fig:b_chi_s}}
\end{figure}

\noindent For a WIMP of spin $j_\chi$ all the operators
$\calO_{X,s,l}$ with $s\le 2 j_\chi$ contribute to the cross
section. Conversely, for a given operator $\calO_{X,s,l}$ the
dependence on $j_\chi$ of the squared amplitude is encoded in the
$B_{j_\chi,s}$ factors. A plot of $B_{j_\chi,s}$ is provided in
Fig.~\ref{fig:b_chi_s}.

Writing:

\be
R_X^{\tau\tau^{\prime}}=R_{0X}^{\tau\tau^{\prime}}+R_{1X}^{\tau\tau^{\prime}} (v^{\perp}_T)^2=R_{0X}^{\tau\tau^{\prime}}+R_{1X}^{\tau\tau^{\prime}}\left (v_T^2-v_{min}^2\right ),
\label{eq:r_decomposition}
\ee
\noindent the correspondence between each of the couplings of
Eq.~(\ref{eq:basis_WIMP_nucleon_operators_alt}) and the values of $X$
in the nuclear response functions
$\widetilde{W}_{TX}^{\tau\tau^{\prime}}$ is provided in
Table~\ref{table:eft_summary}.

\begin{table}[t]
\begin{center}
{\begin{tabular}{|c|c|c||c|c|c|}      
\hline
coupling  &  $R^{\tau \tau^{\prime}}_{0X}$  & $R^{\tau \tau^{\prime}}_{1X}$ & coupling &  $R^{\tau \tau^{\prime}}_{0X}$  & $R^{\tau \tau^{\prime}}_{1X}$ \\
\hline
$M,s,s$  &   $M(q^{2s})$ & - & $\Phi,s,s-1$  &   $\Phi^{\prime\prime}(q^{2s})$,$\tilde{\Phi}^{\prime}(q^{2s})$  & $\Sigma^{\prime}(q^{2s-2})$,$\Sigma^{\prime\prime}(q^{2s-2})$\\
$\Omega,s s$  & -   & $\Sigma^{\prime}(q^{2s})$ & $\Phi,s,s$  &  $\tilde{\Phi}^{\prime}(q^{2s+2})$ & $\Sigma^{\prime}(q^{2s})$ \\
$\Sigma,s,s-1$  & $\Sigma^{\prime\prime}(q^{2s-2})$,$\Sigma^{\prime}(q^{2s-2})$ & - & $\Phi,s,s+1$  & $\Phi^{\prime\prime}(q^{2s+4})$  & $\Sigma^{\prime\prime}(q^{2s+2})$\\
$\Sigma,s,s$  & $\Sigma^{\prime}(q^{2s})$ & - & $\Delta,s,s-1$  &  $\Delta(q^{2s})$  & $M(q^{2s-2})$ \\
$\Sigma,s,s+1$  & $\Sigma^{\prime\prime}(q^{2s+2})$ & - & $\Delta,s,s$  &  $\Delta(q^{2s+2})$  & $M(q^{2s})$ \\
\hline
\end{tabular}}
\caption{The values of $X$ for the nuclear response functions
  $\widetilde{W}_{TX}^{\tau\tau^{\prime}}$ corresponding to each of
  the couplings of Eq.~(\ref{eq:basis_WIMP_nucleon_operators_alt}),
  for the velocity--independent and the velocity--dependent components
  parts of the WIMP response function, decomposed as in
  Eq.~(\ref{eq:r_decomposition}).  In parenthesis the power of $q$ in
  the cross section.
  \label{table:eft_summary}}
\end{center}

\end{table}

\section{Experimental sensitivities to the effective operators}
\label{sec:sensitivities}

\subsection{The scattering rate}
\label{sec:rate}
In this Section we will assume that the WIMP--nucleus scattering
process is driven by each one of the 44 effective couplings in
Eq.~(\ref{eq:basis_WIMP_nucleon_operators_alt}) for $j_\chi\le$ 2, and
obtain constraints from a representative sample of the present DD
experiments by comparing the expected rate in each model to the
corresponding experimental upper bound. In particular the expected
number of events in a WIMP direct detection experiment in the interval
of visible energy $E_1^{\prime}\le E^{\prime}\le E_2^{\prime}$ is
given by:

\begin{equation}
  R_{[E_1^{\prime},E_2^{\prime}]}=\sum_T\int_{E_1^{\prime}}^{E_2^{\prime}}d E^{\prime}\;\left(\frac{dR}{d E^{\prime}} \right)_T
  \label{eq:start1},
\end{equation}

\noindent where $T$ indicates the target nuclei present in the detector.
For a given target $T$: 

\begin{equation}
 \left(\frac{dR}{d E^{\prime}} \right)_T=\int_0^\infty dE_R \left(\frac{dR}{d E_R}\right )_T {\cal
   G}\left [E^{\prime},q_T(E_R)E_R\right ]\epsilon(E^{\prime})
  \label{eq:diff_rate}.
\end{equation}

In the equations above $E_R$ is the recoil energy deposited in the
scattering process (indicated in keVnr), while $E_{ee}$=$q_T(E_R)E_R$
(indicated in keVee) is the fraction of $E_R$ that goes into
ionization and scintillation (wth $q_T(E_R)$ the target quenching
factor) and ${\cal G}(E^{\prime},E_{ee})$ is the effect of the energy
resolution (so that $E^{\prime}$ is measured instead of $E_{ee}$
through a calibration procedure). Finally $\epsilon(E^{\prime})$ is
the measured experimental acceptance. The theoretical differential
rate is given by:

\begin{equation}
\left(\frac{dR}{dE_R}\right)_T=M N_T T_0 \int_{v_{T,min}(E_R)}^{v_{esc}} \frac{\rho_{\chi}}{m_{\chi}}
v \left(\frac{d\sigma}{dE_R}\right)_T\, f(\vec{v}) d^3v,
\label{eq:dr_der}
  \end{equation}

\noindent with $f(\vec{v},t)$ the WIMP velocity distribution in the
Earth's rest frame, while:

\begin{equation}
  v_{T,\rm min}(E_R)=\sqrt{\frac{m_T E_R}{2 \mu_{\chi T}^2}},
  \label{eq:vmin}
\end{equation}

\noindent with $m_T$ the nuclear
target and ${\mu_{\chi T}}$ the WIMP--target reduced mass.

\subsection{Present constraints}
\label{sec:present}

\begin{figure}
\begin{center}
  \includegraphics[width=0.32\textwidth]{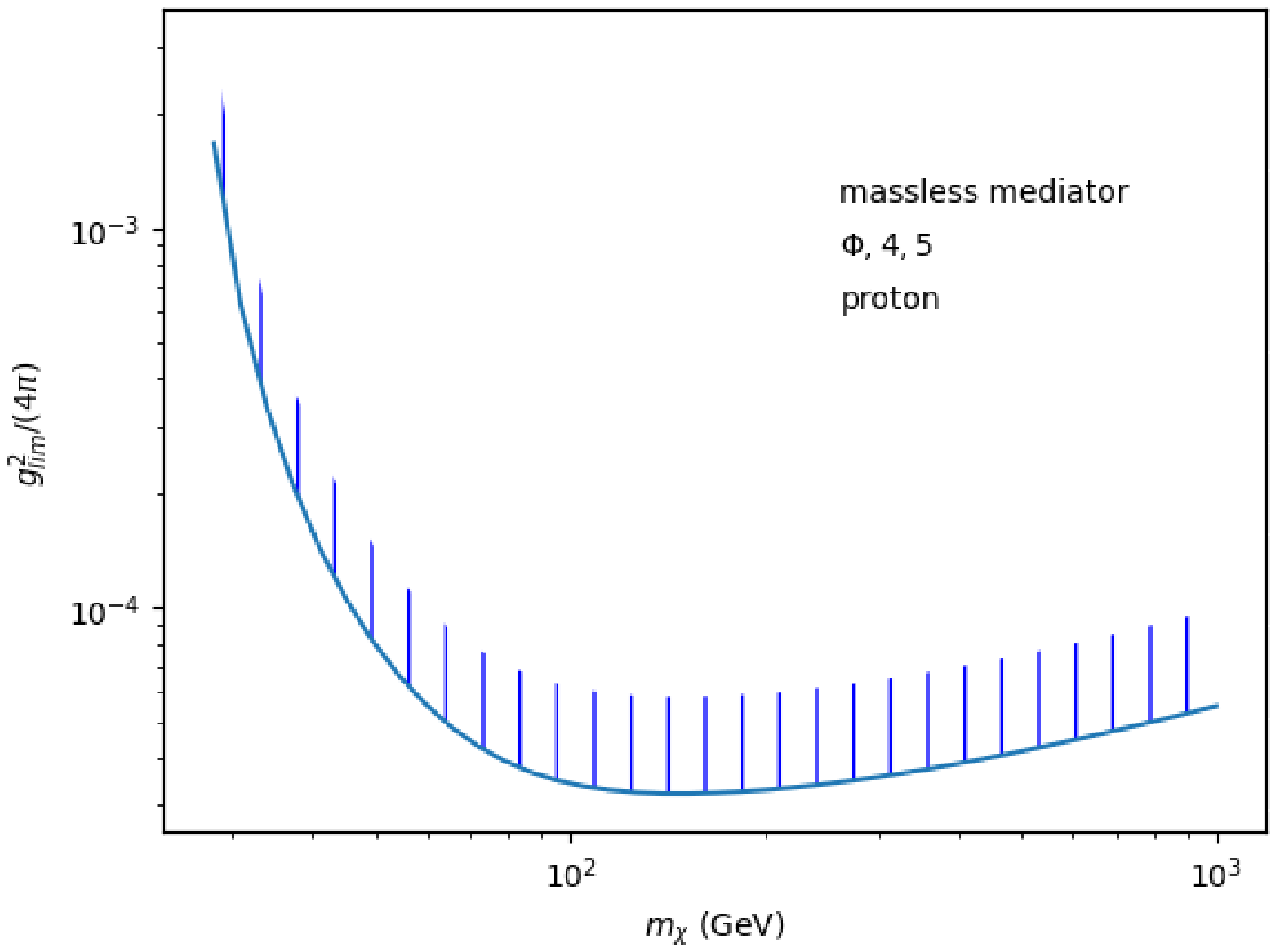}
  \includegraphics[width=0.32\textwidth]{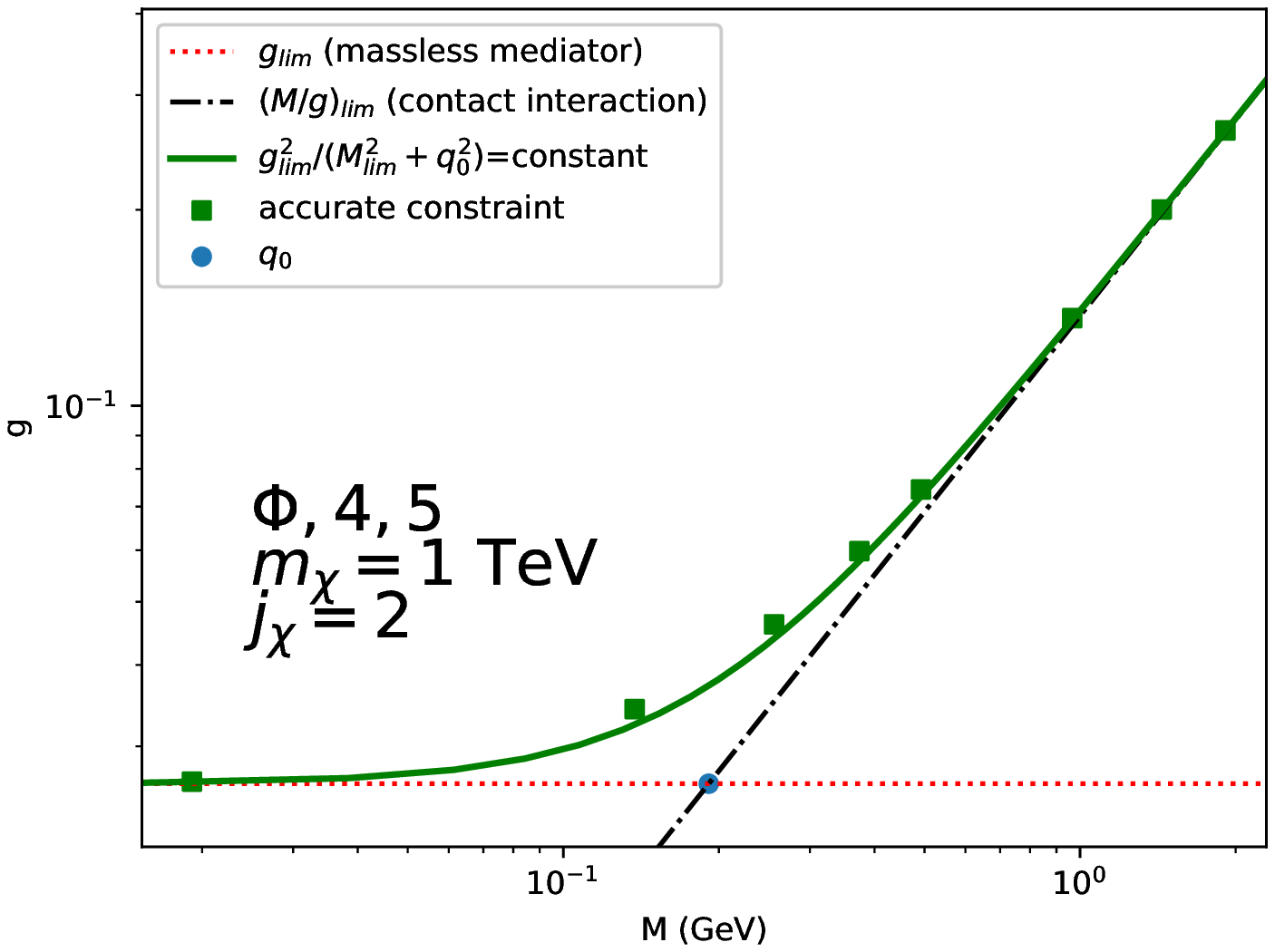}
  \includegraphics[width=0.32\textwidth]{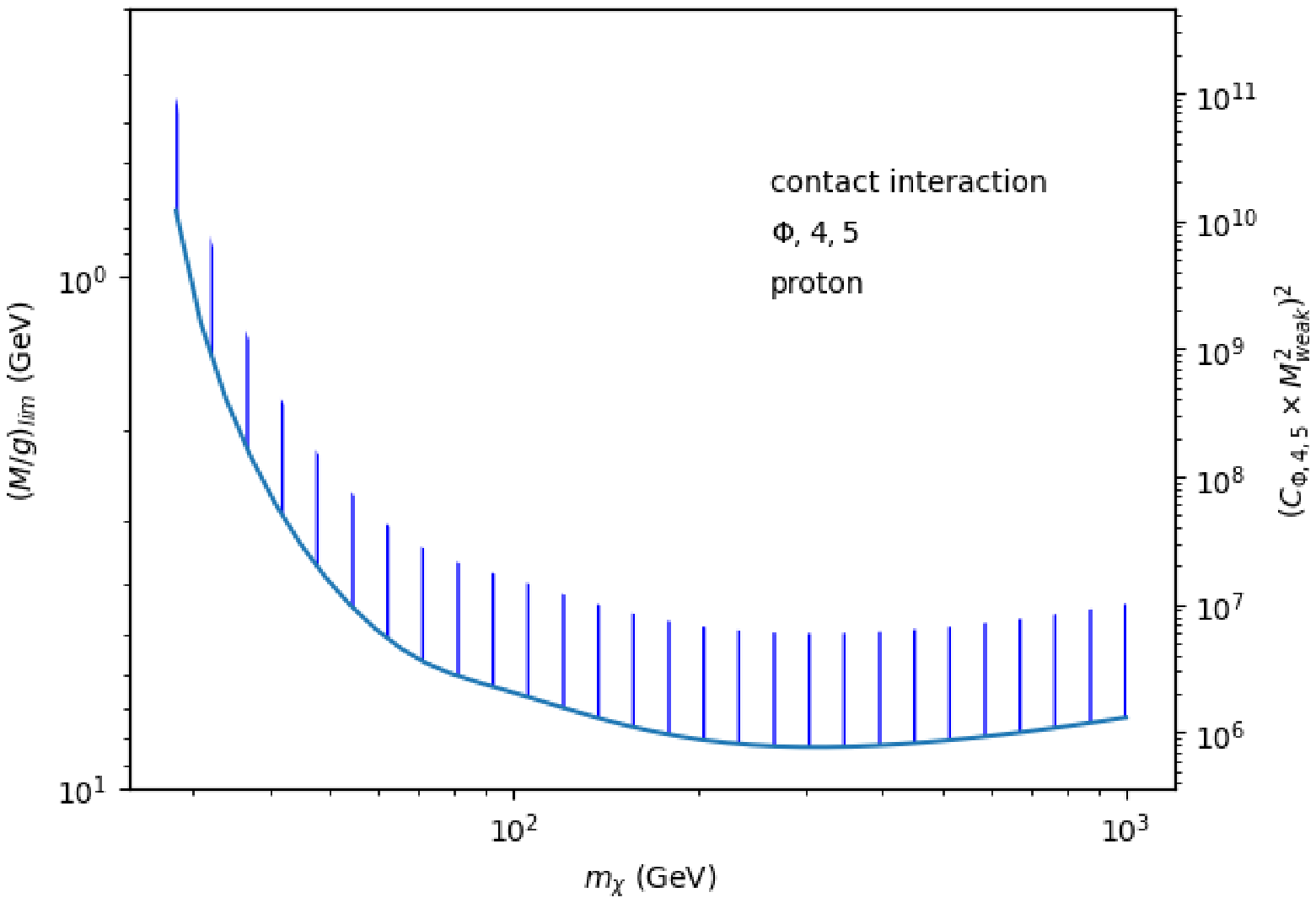}
\end{center}
\caption{Present upper bounds from
  XENON100~\cite{xenon100_high_energy}) on the effective operator
  $\calO_{\Phi,4,5}$ for $j_\chi$=2. {\bf Central plot:} upper bound
  on the coupling $g$ as a function of $M$ for $m_\chi$=1 TeV if the
  Wilson coefficient $c^{\tau}_{\Phi,4,5}(q)$ is parameterized
  according to Eq.~(\ref{eq:c_par}). Square markers: result of an
  accurate evaluation of the bound using the full
  expression~(\ref{eq:c_par}); solid line: curve $g^2/(M^2+q_0^2)$=K,
  where the two constant parameters $q_0$ and $K$ are determined by
  fixing the bounds for $M\ll q$ and $M\gg q$. {\bf Left--hand plot:}
  upper bound on $g$ as a function of $m_\chi$ assuming
  $c^{\tau}_{\Phi,4,5}\simeq g^2/q^2$ (long--range interaction); {\bf
    Right--hand plot:} lower bound on $M/g$ as a function of $m_\chi$
  assuming $c^{\tau}_{\Phi,4,5}\simeq g^2/M^2$ (contact interaction).
\label{fig:long_short_transition}}
\end{figure}

For each of the 44 couplings we parameterize the corresponding Wilson
coefficient as:

\begin{align}
& c^{\tau}_{X,s,l}(q) =\frac{g^2}{M^2+q^2}, 
\label{eq:c_par}
\end{align}

\noindent where $g$ represents an effective coupling constant while
$M$ a mediator mass. In the central plot of
Fig.~\ref{fig:long_short_transition} we show the upper bound on the
coupling $g$ as a function of $M$ for the effective operator
$\calO_{\Phi,4,5}$, $m_\chi$=1 TeV and $j_\chi$=2.  In our analysis we
consider 5 representative direct detection experiments
(XENON1T~\cite{xenon_2018}, XENON100~\cite{xenon100_high_energy},
SuperCDMS~\cite{super_cdms_2017}, PICO--60~\cite{pico60_2019} and
COSINE--100~\cite{cosine_nature} that use 4 different targets: $Xe$,
$Ge$, $C_3F_8$ and $NaI$ (see Appendix~\ref{app:exp} for a summary
of our procedure). In particular, in
Fig.~\ref{fig:long_short_transition} the bound is driven by XENON100,
which provides the stronger constraint.

The central plot of Fig.~\ref{fig:long_short_transition} shows a clear
transition from the regime $M\ll q$, for which
$c^{\tau}_{\Phi,4,5}\simeq g^2/q^2$ (long--range interaction) to
regime for which $M\gg q$ and $c^{\tau}_{\Phi,4,5}\simeq g^2/M^2$
(contact--interaction). In particular the square (green) markers, that
show the result of an accurate evaluation of the bound using the full
expression~(\ref{eq:c_par}), follow closely the solid line that
represents the curve $g^2/(M^2+q_0^2)$=K, where the two parameters
$q_0$ and $K$ are fixed so that the bound for $M\ll q$ coincides to
the constraint $g<g_{lim}$ shown in the left plot for a long--range
interaction (i.e. assuming $c^{\tau}_{\Phi,4,5}\simeq g^2/q^2$) and
for $M\gg q$ coincides to the constraint $M/g<(M/g)_{lim}$ shown in
the right plot for a short--range interaction (i.e. assuming
$c^{\tau}_{\Phi,4,5}\simeq g^2/M^2$). The transition between the two
asymptotic regimes, represented in the central plot of
Fig.~\ref{fig:long_short_transition} by the horizontal (red) dotted
line and by the dot--dashed (black) line corresponds to $M$ = $q_0$ =
$g_{lim} (M/g)_{lim}$.

\begin{figure}
\begin{center}
  \includegraphics[width=0.49\textwidth]{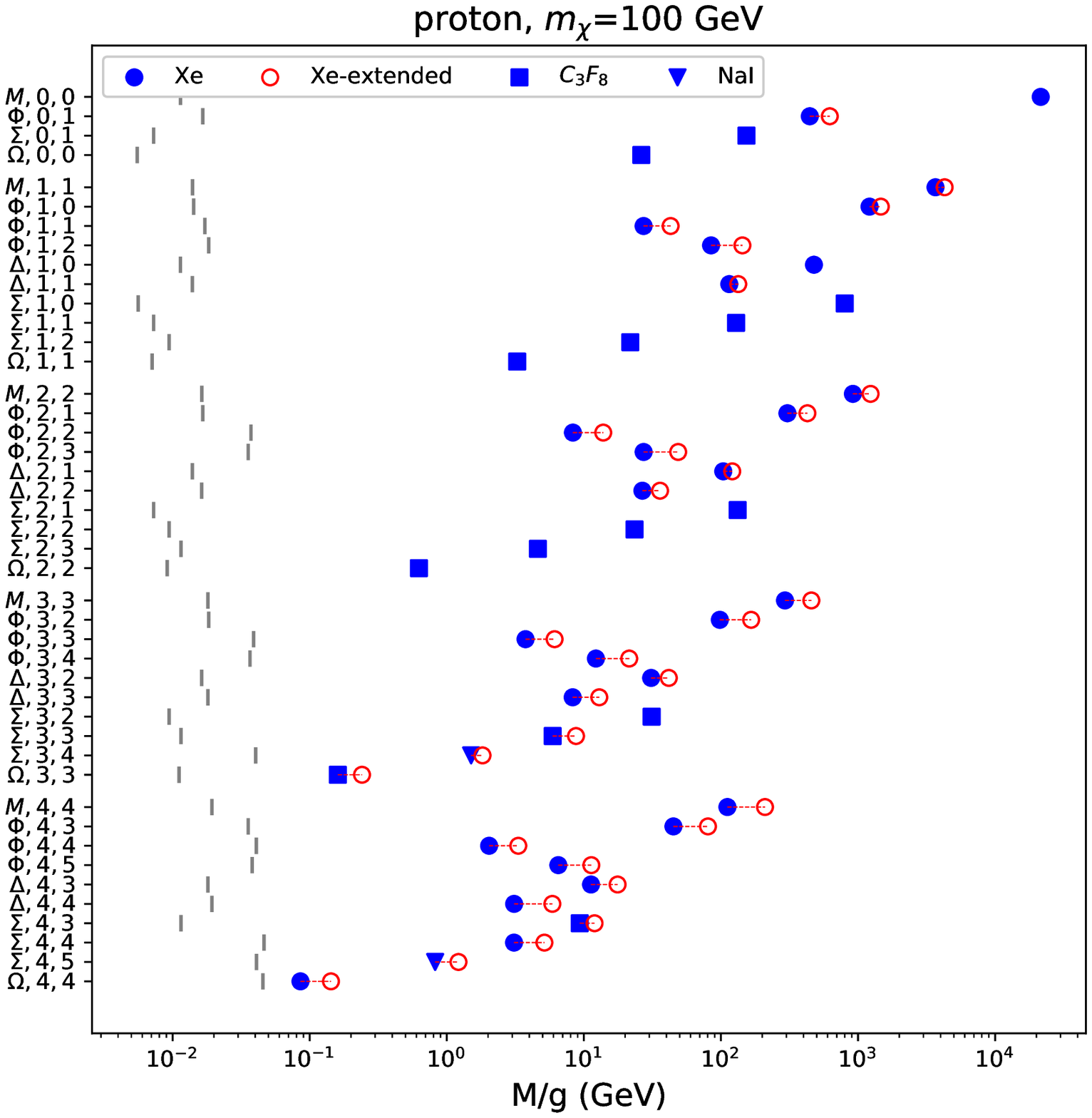}
  \includegraphics[width=0.49\textwidth]{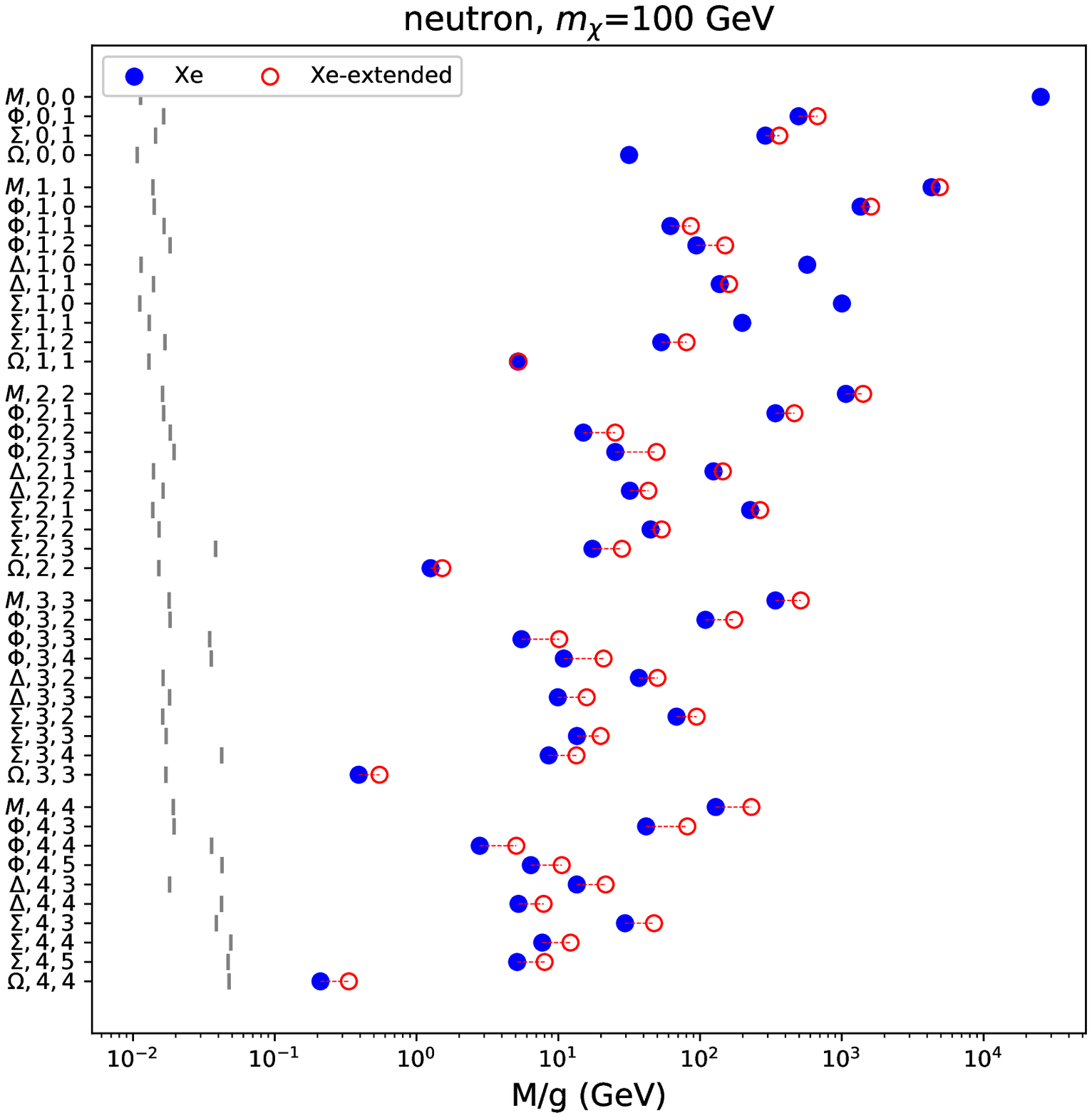}
\end{center}
\caption{Most constraining lower bound on $(M/g)$ for $m_\chi$=100 GeV
  and for all the 44 operators of
  Eq.~(\ref{eq:basis_WIMP_nucleon_operators_alt}) with $j_\chi\le$2,
  and assuming a contact interaction ($c^{\tau}_{X,s,l}= g^2/M^2$ in
  Eq.~(\ref{eq:c_par})). {\bf Left--hand plot:} WIMP--proton
  interaction; {\bf Right--hand plot:} WIMP--neutron interaction. For
  each operator $\calO_{X,s,l}$ the bound assumes $j_\chi=s/2$. Filled
  markers: most constraining present bound from one among the
  experiments analyzed in Appendix~\ref{app:exp}.  Open markers:
  estimation of the improvement on the limits by extending the
  experimental energy ranges beyond the present ones, as explained in
  Section~\ref{sec:limit_improvements}. Vertical solid lines: minimal
  value of $M/g$ compatible with the assumption of a contact
  interaction.
\label{fig:markers_plot_mchi_100}}
\end{figure}

\begin{figure}
\begin{center}
  \includegraphics[width=0.49\textwidth]{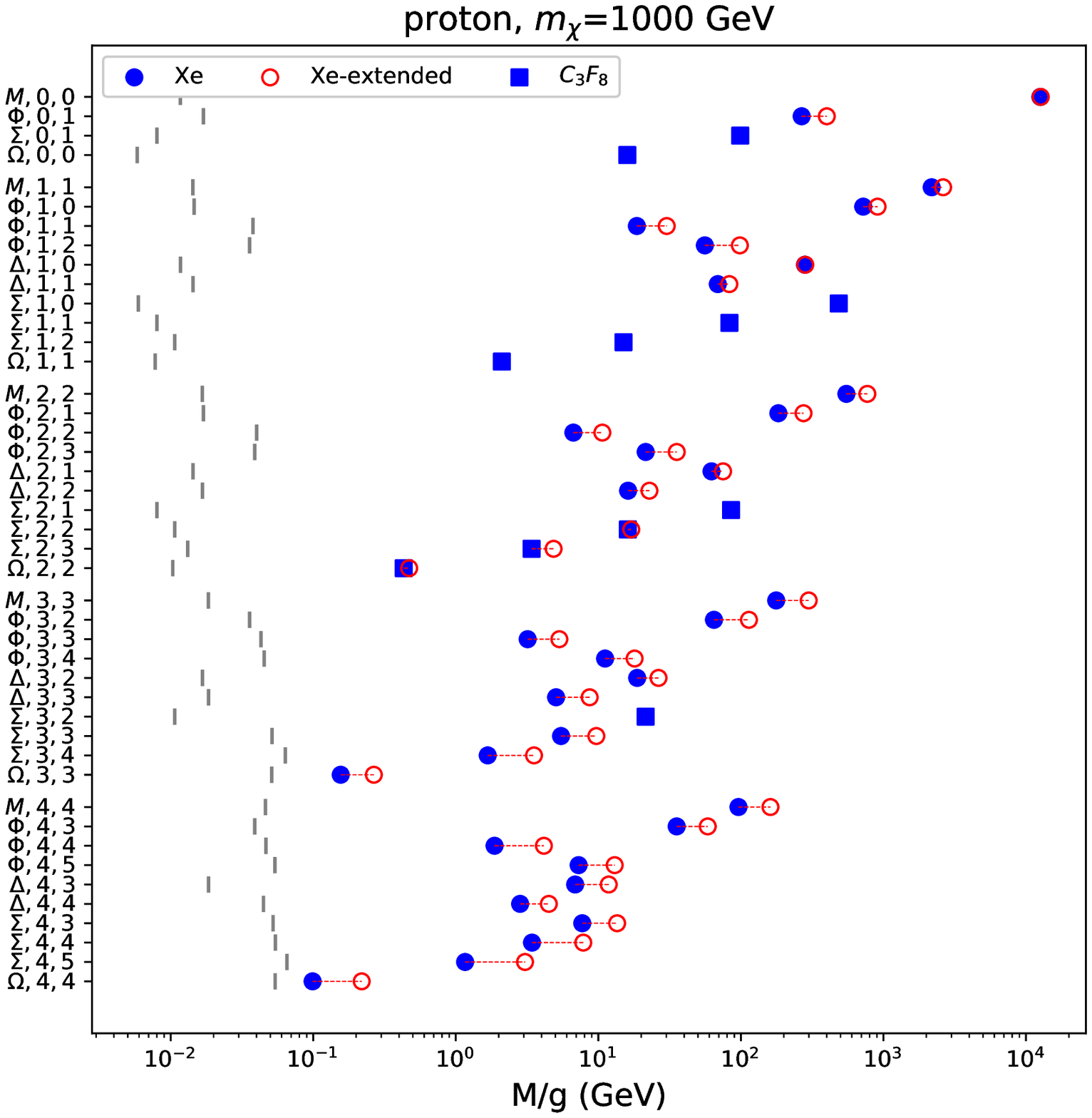}
  \includegraphics[width=0.49\textwidth]{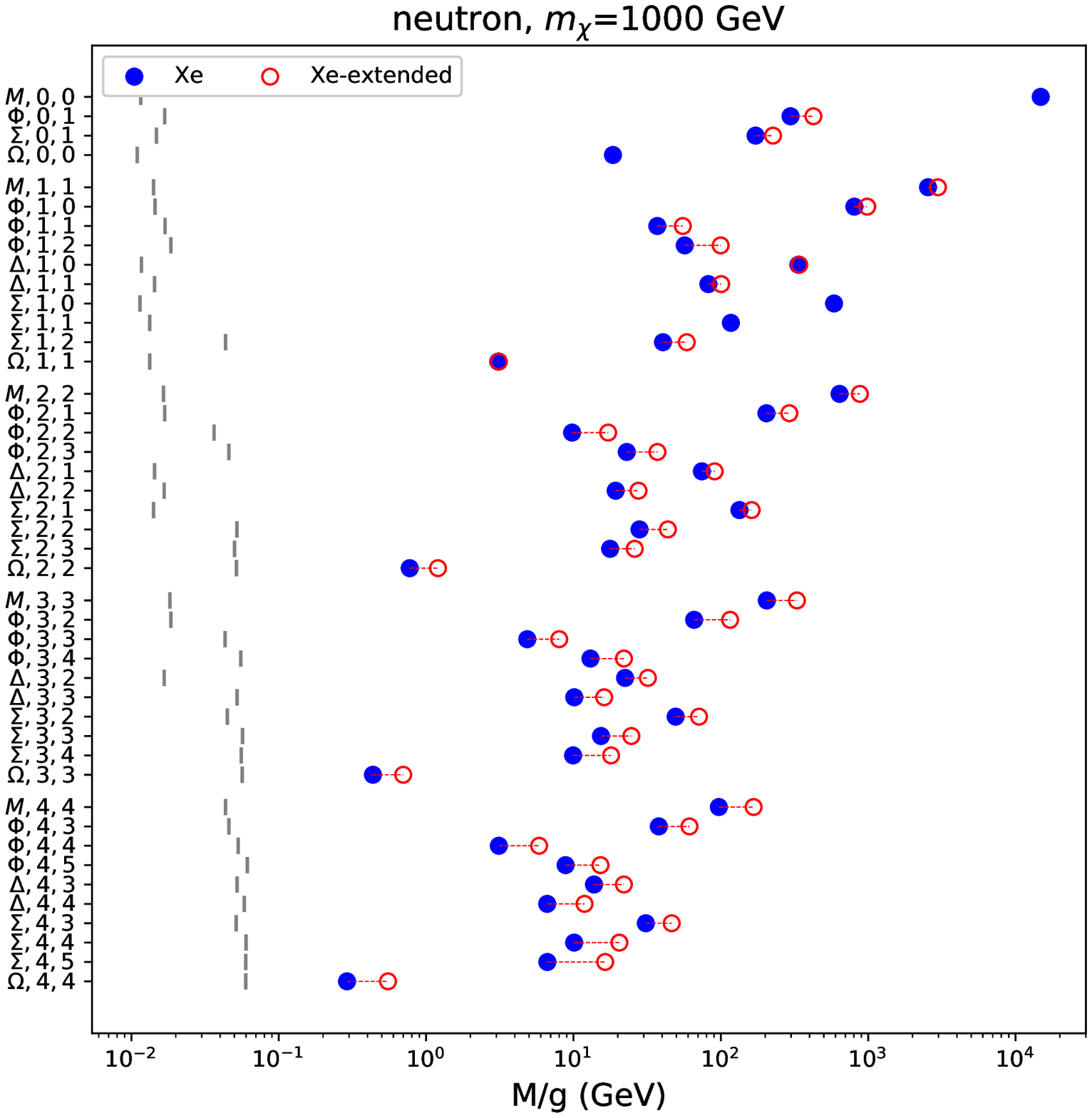}
\end{center}
\caption{Same as in Fig.~\ref{fig:markers_plot_mchi_100} for $m_\chi$=1 TeV.
\label{fig:markers_plot_mchi_1000}}
\end{figure}

In Figs.~\ref{fig:markers_plot_mchi_100}
and~\ref{fig:markers_plot_mchi_1000} we show the result of a
systematic analysis on the most constraining lower bound $(M/g)_{lim}$
for $m_\chi$=100 GeV and $m_\chi$=1 TeV, respectively for all the 44
models and for a contact interaction ($c^{\tau}_{X,s,l}= g^2/M^2$). In
each figure the left--hand plot assumes that the WIMP couples to
protons only, and the right--hand plot to neutrons only. For each
operator $\calO_{X,s,l}$ the bound assumes $j_\chi$=s/2. The
constraint for $j_\chi>$s/2 can be obtained from that for $j_\chi$ =
s/2 by multiplication times the factor
$(B_{s/2,s}/B_{j_\chi,s})^{1/4}$ (with $B_{j_\chi,s}$ given in
Eq.~(\ref{eq:b_jchi_s}) and plotted in Fig.~\ref{fig:b_chi_s}).  In
all the figures each filled marker represents the most constraining
present bound from one among the five experiments that we analyze
while, if present, an open marker indicates an estimation of the
improvement that could be obtained in the limit by extending the
experimental energy ranges beyond the present ones (see
Section~\ref{sec:limit_improvements}). Moreover, for each effective
model, a vertical line represents the minimal value of $M/g$
compatible to the assumption of a contact interaction, obtained by
combining $M>q_0$ (following for each model the procedure outlined in
Fig.~\ref{fig:long_short_transition}) and the perturbativity
requirement $g^2/(4\pi)<$1.

A first conclusion one can draw from
Figs.~\ref{fig:markers_plot_mchi_100}
and~\ref{fig:markers_plot_mchi_1000} is that for all the couplings the
bound on $M/g$ is compatible to the assumption of a contact
interaction. Moreover, one observes an anticorrelation between the
value of $(M/g)_{lim}$ and the $s$ parameter. This is due to the fact
that a larger value of $s$ corresponds to stronger momentum
suppression in the expected rate and, as a consequence, to a weaker
bound. More specifically, as shown in Table~\ref{table:eft_summary}
for a given value of $s$ the power of the transferred momentum $q$
ranges from $2s-2$ and $2s+4$.

In
Figs.~\ref{fig:markers_plot_mchi_100}--~\ref{fig:markers_plot_mchi_1000}
the bounds on $M/g$ span about 5 orders of magnitude, ranging from
$\sim$ 10--20 TeV down to $\sim$ 90--200 MeV, The strongest
constraint always corresponding to the operator $\calO_{M,0,0}$ and
the weaker constraint corresponding to the operator
$\calO_{\Omega,4,4}$.  This hierarchy is not surprising, since
$\calO_{M,0,0}$ corresponds to the standard spin--independent
interaction (scaling with the square of the nuclear mass number) with
no momentum or velocity suppression, while $\calO_{\Omega,4,4}$ is the
operator with the highest momentum suppression in our analysis
(i.e. for $j_ \chi\le$ 2) among the velocity--suppressed ones
(i.e. with $R_{0X}^{\tau\tau'}$=0 in Eq.~(\ref{eq:r_decomposition})).
Moreover, with the exception of the case $s$ = 0 for which only 4
operators are defined, for each of the other values of $s$ the
corresponding constraints on the $M/g$ for the corresponding 10
operators appear to present a similar structure. Focusing for instance
on $s$=1, the bound for $\calO_{M,1,1}$ is the most constraining, with
$\calO_{\Phi,1,0}$ coming next in order of size. This can be
understood by noticing that, as shown in
Table~\ref{table:eft_summary}, the cross section for
$\calO_{\Phi,0,1}$ has the same momentum suppression as
$\calO_{M,1,1}$ and depends on the nuclear response function
$W_{\Phi^{\prime\prime}}^{\tau\tau^{\prime}}$.  Such nuclear response
function favors heavier elements with large nuclear shell and scales
with the nuclear target similarly to the SI interaction, so that for
most isotopes it is the most sizeable among the
$W_{TX}^{\tau\tau^{\prime}}$ with the exception of
$X=M$~\cite{haxton1,haxton2} (this holds for all the nuclear targets
that we consider with the exception of the semi-magic isotope
$^{72}Ge$, for which $W_{\Phi^{\prime\prime}}^{\tau\tau^{\prime}}$
vanishes). As far as the other operators are concerned, for $X$=
$\Delta$, $\Sigma$ the constraint gets systematically less stringent
at growing $l$, as one expects due to the enhanced momentum
suppression, i.e. $(M/g)_{lim}(\Delta,1,0) > (M/g)_{lim}(\Delta,1,1) >
(M/g)_{lim}(\Delta,1,2)$ and $(M/g)_{lim}(\Sigma,1,0) >
(M/g)_{lim}(\Sigma,1,1) > (M/g)_{lim}(\Sigma,1,2)$ An exception to
this pattern is provided by $X$ = $\Phi$ for which one observes
instead $(M/g)_{lim}(\Phi,1,0) > (M/g)_{lim}(\Phi,1,2) >
(M/g)_{lim}(\Phi,1,1)$. One understands this inversion between
$(M/g)_{lim}(\Phi,1,1)$ and $(M/g)_{lim}(\Phi,1,2)$ by noticing that,
as again shown in Table~\ref{table:eft_summary}, $\calO_{\Phi,1,1}$
couples to $W_{\tilde{\Phi}^{\prime}}^{\tau\tau^{\prime}}(q)$ while
$\calO_{\Phi,1,2}$ and $\calO_{\Phi,1,0}$ couple to
$W_{\Phi^{\prime\prime}}^{\tau\tau^{\prime}}(q)$, with
$W_{\Phi^{\prime\prime}}^{\tau\tau^{\prime}}(\q)\gg
W_{\tilde{\Phi}^{\prime}}^{\tau\tau^{\prime}}(\q)$, as already pointed
out. The same pattern among the 10 corresponding constraints on $M/g$
repeats also for $s$ =2,3,4.

Another feature arising from Figs.~\ref{fig:markers_plot_mchi_100}
and~\ref{fig:markers_plot_mchi_1000} is that in most cases the
strongest constraint is provided by xenon detectors (XENON1T and
XENON100) with the few exceptions of fluorine (PICO--60) and $NaI$
(COSINE--100), but only for a WIMP--proton interaction.

Indeed in most cases xenon experiments are the most competitive due to
the large exposure.  In order to understand why there are exceptions
to this one needs again to consider the mapping provided in
Table~\ref{table:eft_summary} between each effective coupling and the
nuclear response functions $W^{\tau\tau^{\prime}}_X$. In particular
the 10 operators arising at a given value of $s$ can be divided into 4
broad classes: i) operators whose cross section is driven by the
$W^{\tau\tau^{\prime}}_M$ response function: $\calO_{M,s,s}$,
$\calO_{\Delta,s,s-1}$ and $\calO_{\Delta,s,s}$; ii) operators driven
by either the $W^{\tau\tau^{\prime}}_{\Sigma^{\prime\prime}}$ or the
$W^{\tau\tau^{\prime}}_{\Sigma^{\prime}}$ response function:
$\calO_{\Sigma,s,-1}$, $\calO_{\Sigma,s,s}$, $\calO_{\Sigma,s,s+1}$
and $\calO_{\Omega,s,s}$; iii) operators driven by the
$W^{\tau\tau^{\prime}}_{\Phi^{\prime\prime}}$ response function:
$\calO_{\Phi,s,s-1}$ and $\calO_{\Phi,s,s+1}$ iv) the operator
$\calO_{\Phi,s,s}$, which is driven by
$W^{\tau\tau^{\prime}}_{\tilde{\Phi}^{\prime}}$.  In particular we
notice that the operators $\calO_{\Delta,s,s-1}$ and
$\calO_{\Delta,s,s}$ couple to $W^{\tau\tau^{\prime}}_M$ through a
velocity--suppressed term (i.e. $R_{1X}^{\tau\tau^{\prime}}$ in
Eq.~(\ref{eq:r_decomposition})). However, in spite of the $v_T^2/c^2
\simeq 10^{-6}$ suppression, for heavy nuclei ($Xe$, $I$) the
scattering amplitude of such operators is dominated by the
$R_{1X}^{\tau\tau^{\prime}}$ terms thanks to the huge hierarchy
between the two response functions $W^{\tau\tau^{\prime}}_M \gg
W^{\tau\tau^{\prime}}_\Delta$.  On the other hand, for lighter targets
($Ge$, $Na$, $F$) the contributions from $R_{0X}^{\tau\tau^{\prime}}$
and $R_{1X}^{\tau\tau^{\prime}}$ are of the same order.  This feature
was already observed in the context of spin--1/2 WIMPs for $\calO_5$
=- $\calO_{\Delta,1,1}$ and $\calO_8$ =
$\calO_{\Delta,1,0}$~\cite{sogang_scaling_law_nr}.

The mapping summarized above allows to understand that xenon
experiments are the most sensitive to the couplings driven by
$W^{\tau\tau^{\prime}}_M$ and
$W^{\tau\tau^{\prime}}_{\Phi^{\prime\prime}}$, which favor heavy
elements. On the other hand,
$W^{\tau\tau^{\prime}}_{\Sigma^{\prime\prime}}$ and
$W^{\tau\tau^{\prime}}_{\Sigma^{\prime}}$ correspond to a coupling of
the WIMP to the nucleon spins.  Since inside nuclei the nucleon spins
tend to cancel each other the contribution from even-numbered nucleons
to such response functions is strongly suppressed. As a consequence of
this for such interactions neutron-odd targets (such as xenon and
germanium) are mostly sensitive to the WIMP--neutron coupling, while
proton--odd targets (such as fluorine, sodium and iodine) are mostly
sensitive to the WIMP--proton coupling. This implies that in the
left--hand plots of Figs.~\ref{fig:markers_plot_mchi_100}
and~\ref{fig:markers_plot_mchi_1000} the rate in xenon detectors is
strongly suppressed for $X$ = $\Sigma$, $\Omega$. In such cases the
strongest constraint is provided by fluorine in PICO--60 which, due to
the large exposure, is the most competitive among proton--odd
detectors, albeit only for moderate values of $s$. Indeed, for $s$ =
3, 4 the strongest constraint can instead be provided by xenon or
iodine.  We postpone the explanation of this fact to
Section~\ref{sec:diff_rate}, where the expected spectral shape of the
differential rate in our scenarios will be discussed.  Finally, the
response function $W^{\tau\tau^{\prime}}_{\tilde{\Phi}^{\prime}}$
requires a nuclear spin $\ge 1$~\cite{haxton1} and vanishes in
fluorine, so that also in this case xenon turns out to be the most
competitive target.

The results shown in Figs.~\ref{fig:long_short_transition},
\ref{fig:markers_plot_mchi_100} and \ref{fig:markers_plot_mchi_1000}
are also valid in the case of the multipolar DM discussed in
Section~\ref{sec:high_multipole}.  Notice that the interaction driving
the scattering process and the binding force within the DM composite
state need not be the same, and that the latter is not accessible
experimentally. If, on the other hand, one assumes the same
interaction, and that during the scattering process polarization
does not induce lower multipoles in the DM state, the
nonrenormalizable operators in the multipolar effective Lagrangian
depend on the particle mass or the particle radius and not on the
energy cutoff scale at which the interactions become weak or strong,
or at which new degrees of freedom become dynamical. For instance, the
analysis in~\cite{porrati_rahman} concludes that the generic energy
cutoff for effective theories of electromagnetic interactions of a
particle with mass $m$, spin $j$ and coupling $g$ is $\sim
m/g^{1/(2j-1)}$ for $j\ge$ 3/2.  

We conclude this Section by pointing out that within the context of
the non--relativistic effective theory used in our analysis it is not
possible to assess the compatibility of the bounds obtained in
Figs.~\ref{fig:markers_plot_mchi_100} and
\ref{fig:markers_plot_mchi_1000} for the mediator mass $M$ with other
constraints, such as those from accelerator physics. Instead, in the
next Section we will use such bounds to calculate the expected maximal
signals in future direct detection experiments.

\begin{figure}
  \begin{center}
    \includegraphics[width=11cm]{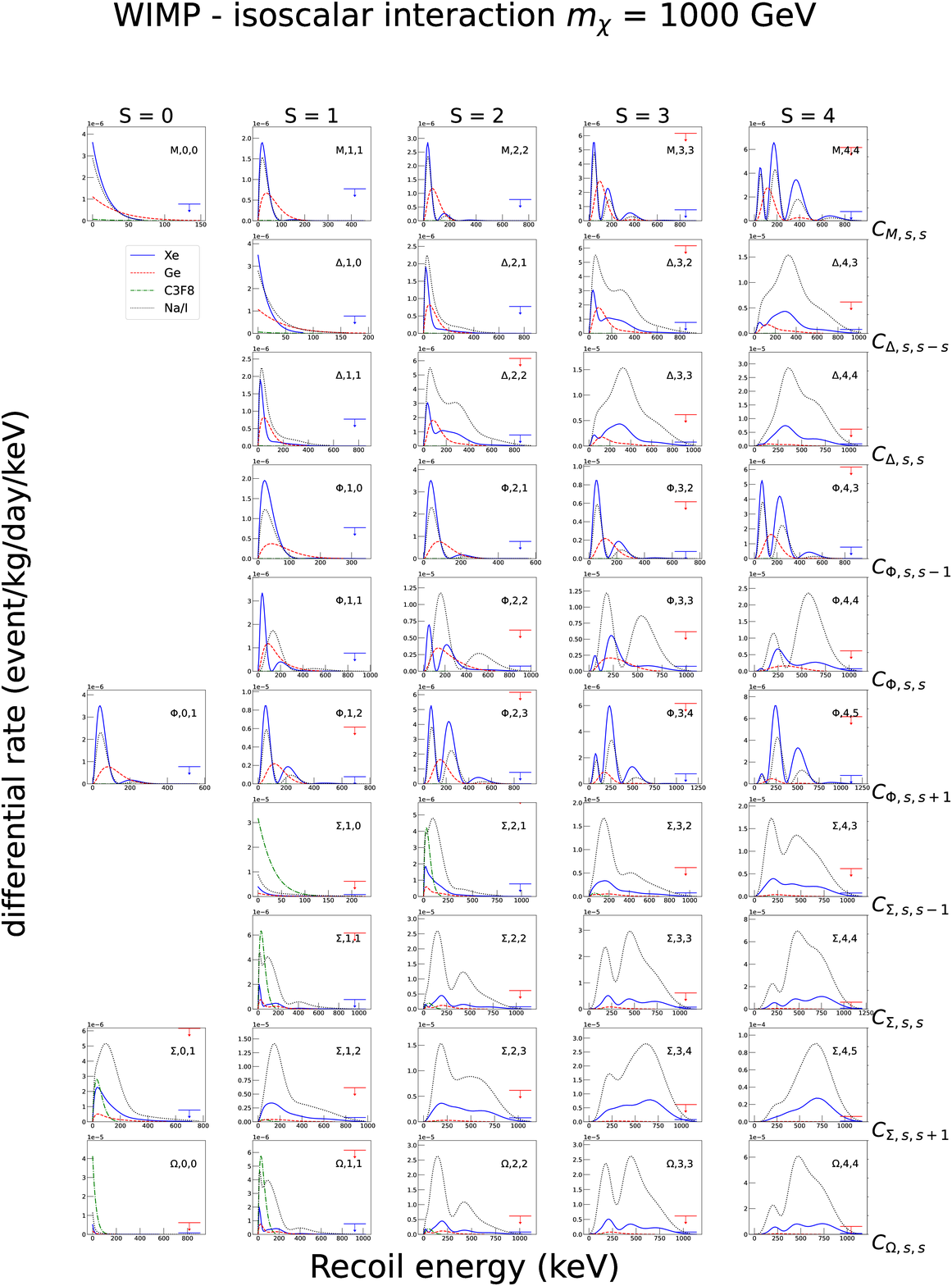}
    \end{center}
  \caption{Differential rate as given by Eq.~(\ref{eq:diff_rate}) as a
    function of the nuclear recoil energy $E_R$ for each of the 44
    couplings in Eq.~(\ref{eq:basis_WIMP_nucleon_operators_alt}) for
    $m_\chi$=1 TeV and an isoscalar interaction ($c^{p}_{X,s,l}$ =
    $c^{n}_{X,s,l}$). In each plot $j_\chi$=s/2, while $M/g$ is fixed
    to the present upper bound. The horizontal lines show the residual
    background levels achieved by present experiments in the
    low--energy part of the spectrum.
\label{fig:poster_plot}}      
\end{figure}

\subsection{Energy spectra for high--spin WIMPs}
\label{sec:diff_rate}

Besides leading to a suppression of the expected rates, the larger
powers of the momentum transfer $q$ = $\sqrt{2 m_T E_R}$ that arise in
high--multipole operators lead to another important effect: they push
the expected differential spectra to larger recoil energies $E_R$.  An
example of this is shown in Fig~\ref{fig:poster_plot}, where the
differential rate of Eq.~(\ref{eq:diff_rate}) is plotted for each of
the 44 couplings fixing $m_\chi$=1 TeV, assuming an isoscalar
interaction ($c^{p}_{X,s,l}$ = $c^{n}_{X,s,l}$) and fixing $M/g$ to
the present upper bound, calculated as in
Figs.~\ref{fig:markers_plot_mchi_100}
and~\ref{fig:markers_plot_mchi_1000}.

\begin{figure}
\begin{center}
  \includegraphics[width=0.49\textwidth]{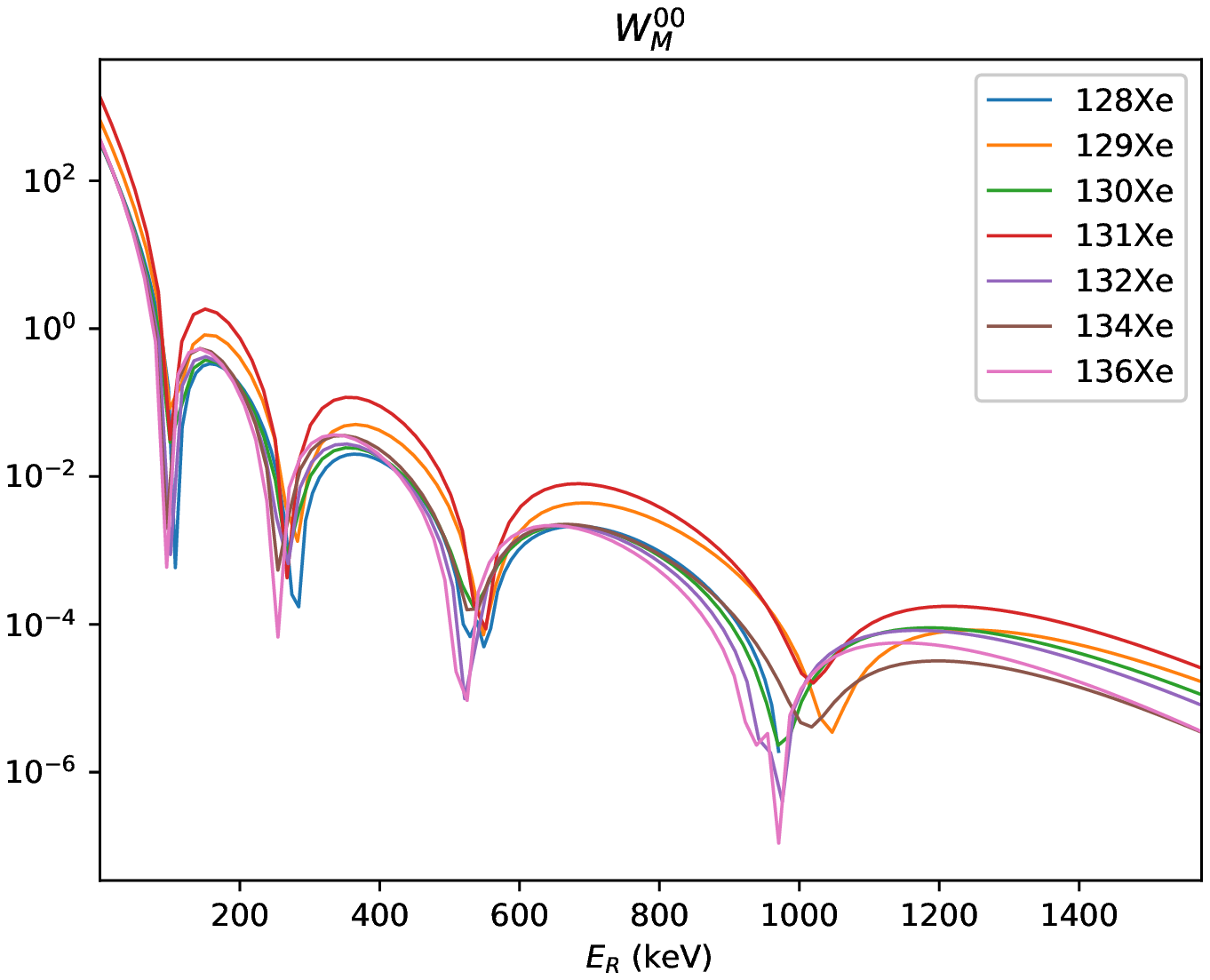}
  \includegraphics[width=0.49\textwidth]{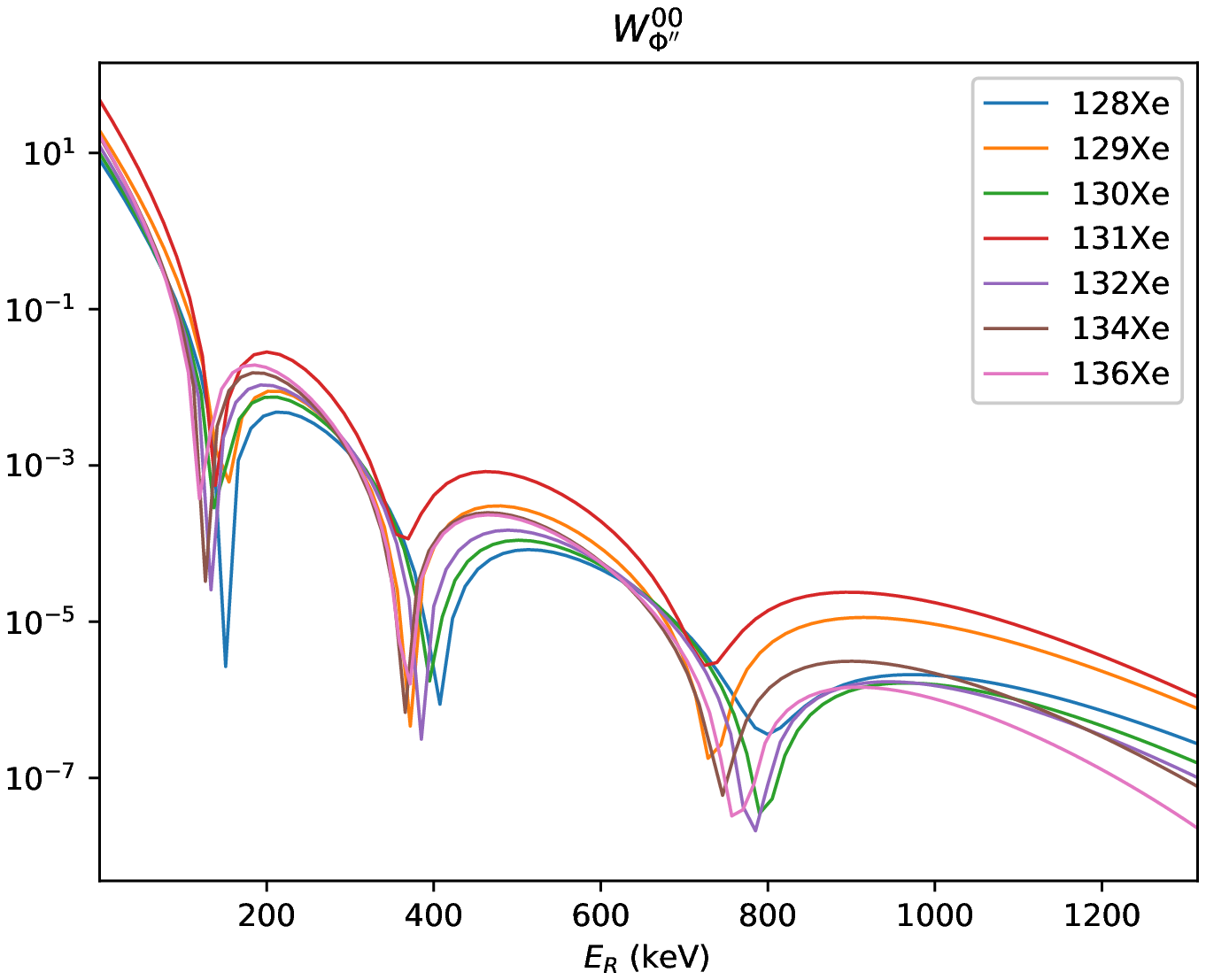}
\end{center}
\caption{Nuclear structure functions $W^{00}_M$ and
  $W^{00}_{\Phi^{\prime\prime}}$ for xenon as calculated
  in~\cite{haxton2} as a function of the nuclear recoil energy $E_R$ for
  an isoscalar interaction ($\tau$ = $\tau^{\prime}$ = 0).
  \label{fig:w_func_xe}}
\end{figure}

\begin{figure}
\begin{center}
  \includegraphics[width=0.99\textwidth, height=0.2\textheight]{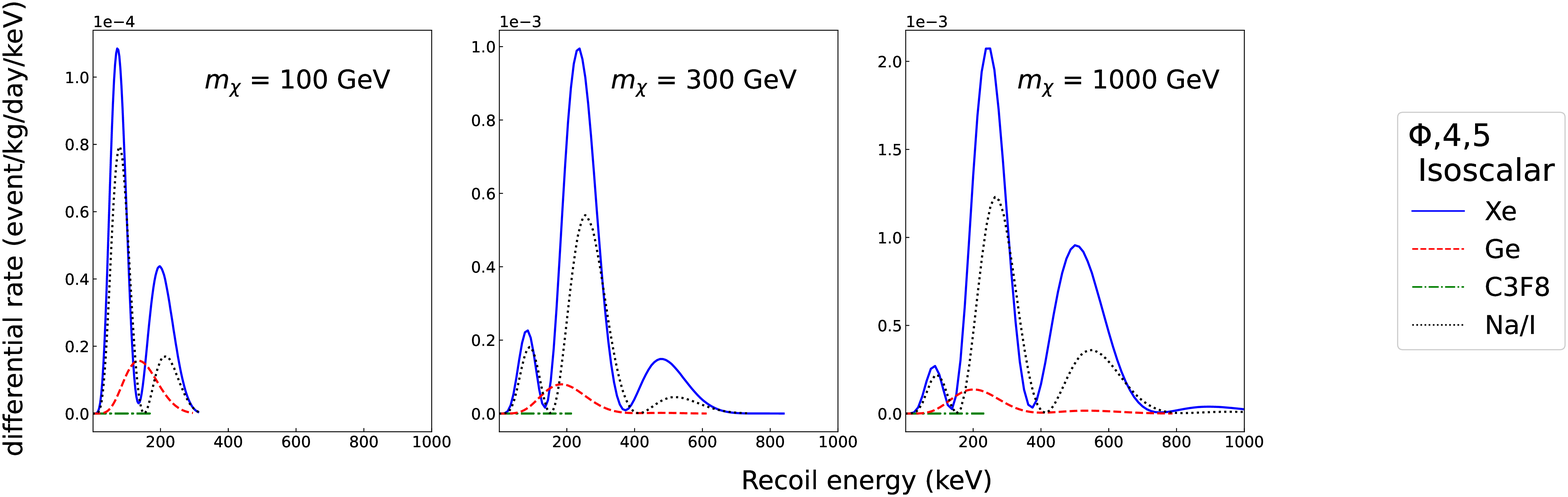}
\end{center}
\caption{Expected differential rates for $\calO_{\Phi,4,5}$ calculated
  with the same assumptions of Fig.~\ref{fig:poster_plot}. {\bf
    Left--hand plot:} $m_\chi$=100 GeV; {\bf
    Central plot:} $m_\chi$=300 GeV; {\bf Right--hand plot:} $m_\chi$=1 TeV.
\label{fig:peak_change}}
\end{figure}

Indeed, in some cases the WIMP signal not only extends well beyond
$E_R$ = 1 MeV (for high multipoles $s$ and for large--enough values of
the WIMP mass $m_\chi$ and of the target mass $m_T$) but its largest
part is concentrated to $E_R\gsim$ 100 keV, showing there a structure
of peaks and minima.  The peculiar features of such plots can be
understood as a combination of three ingredients: the nuclear
structure function $W^{\tau\tau^{\prime}}_X$, the WIMP velocity
distribution $f(\vec{v})$ and the power of $q$ that appears in the
cross section of each operator (as summarized in
Table~\ref{table:eft_summary}).  In particular the peaks appearing in
Fig.~\ref{fig:poster_plot} descend directly from the diffractive
nature of the $W^{\tau\tau^{\prime}}_{TX}$ nuclear structure response
functions, which are calculated as Fourier transforms of nuclear
current densities. An example of the $W^{\tau\tau^{\prime}}_M$ and
$W^{\tau\tau^{\prime}}_{\Phi^{\prime\prime}}$ functions for xenon as
calculated in~\cite{haxton2} is shown in Fig.~\ref{fig:w_func_xe} for
an isoscalar interaction ($\tau$ = $\tau^{\prime}$ = 0).  In the
differential rate the $W^{\tau\tau^{\prime}}_{TX}$ function is
convoluted with the velocity distribution $f(\vec{v})$, which in the
standard halo model we adopt is exponentially suppressed when
$v_{T,\rm min}(E_R)$ approaches the escape velocity. For a standard SI
or SD interaction with no explicit momentum dependence such
suppression prevents the peak structure of the
$W^{\tau\tau^{\prime}}_{TX}$ to emerge, and leads to an energy
spectrum that falls monotonically with the recoil energy. However, in
the general case of the $\calO_{X,s,l}$ operators when $s$ is large
enough the power of $q$ appearing in the cross section enhances the
peaks at high recoil energies, so that they become visible, if, at the
same energies, the spectrum is not cut by the velocity distribution.
For heavy nuclei ($Xe$, $I$) this is indeed the case, so that at large
$s$ the diffractive peaks become visible (notice that, in the most
extreme case, the cross section for $\calO_{\Phi,4,5}$ is enhanced at
high recoil energy by a $q^{12}\simeq E_R^6$ factor).  This effect is
also enhanced for heavier WIMP masses, as illustrated in
Fig.~\ref{fig:peak_change}, where the expected differential rates,
calculated with the same assumptions of Fig.~\ref{fig:poster_plot},
are shown for $\calO_{\Phi,4,5}$ and $m_\chi$ = 100 GeV, 300 GeV and 1
TeV: in particular as $m_\chi$ is increased the overall signal rate at
high energy gets larger while the spectrum becomes harder, with a
growing relative size of the peak with the highest energy. A different
situation is observed however for lighter nuclei ($F$, $Ge$, $Na$). In
this case, irrespective on $m_\chi$, the escape velocity cut on the
recoil energy is in the 100--200 keV range, and the diffractive peaks
correspond to values of $v_T$ beyond the escape velocity. As a
consequence, as far as the overall size of the expected signal for
$\calO_{X,s,l}$ is concerned, larger $l$ values (i.e. larger powers of
$q$ in the operator) increasingly favour larger--mass targets compared
to lighter ones. Another way to see this is that, due to the cut from
the escape velocity, the average momentum transfer $q_{light}$ of the
expected spectrum of a light target is smaller than the corresponding
$q_{heavy}$ of the spectrum of a heavy target, so that the overall
rate of a heavy target is enhanced compared to that of a light one by
a factor $(q_{heavy}/q_{light})^n\gg$ 1 if the scattering amplitudes
depend on $q^n$ with $n\gg $1. This effect is observed in the
left--hand plots of Figs.~\ref{fig:markers_plot_mchi_100}
and~\ref{fig:markers_plot_mchi_1000}, where large--enough values of
$l$ suppress the sensitivity of $F$ in favour of the heavier $Xe$ and
$I$ targets, that pick up and become the more competitive ones. In
particular the latter turn out to have a comparable sensitivity
because $Xe$ is neutron odd and is disfavoured by the coupling,
although it is favoured by the large exposure and low residual
background of the XENON1T and XENON100 experiments, while $I$ is
proton--odd, so is favoured by the coupling, although it is
disfavoured by the much lower exposure and the higher residual
background of the COSINE-100 experiment.

In light of the above discussion, heavy targets are the most suitable
to search for large--spin WIMPs, if their expected rate is driven by a
large-multipole operator~\footnote{For this reason also Tungsten (W)
  in CRESST~\cite{cresst_II} would be an excellent target to search
  for high--spin WIMPs. We did not include CRESST in our analysis
  because the nuclear response functions $W_{TX}^{\tau\tau^{\prime}}$
  for tungsten are not available in the literature.}. The high recoil
energy behaviour of the spectra in Fig.~\ref{fig:poster_plot} is at
strong variance with the usual WIMP DD paradigm, where the search for
new physics is focused on the lowest--energy part of the spectrum.
Indeed, among present experiments only PICO--60, that is a threshold
detector, includes the full range of recoil energies in its analysis
(although, as already explained, $F$ targets are only sensitive to
$E_R\lsim$ 100--200 keVnr because of the cut from the velocity
distribution). For instance, the Region of Interest (ROI) analyzed by
XENON1T~\cite{xenon_2018} is limited to $E_R\lsim$ 27 keV, that of
SuperCDMS~\cite{super_cdms_2017} stops at $E_R=$ 100 keV, while that
from COSINE-100~\cite{cosine_nature} is limited to $E_{ee}\le$ 6
keVee, corresponding to $E_R\le$ 18 keVnr for sodium targets and
$E_R\le$ 60 keVnr for iodine targets. Clearly, all such experimental
searches are not optimized to look for some of the signals shown in
Fig.~\ref{fig:poster_plot}.  In addition, COSINE-100 achieved an
exclusion plot at the level of the DAMA effect (using the same $NaI$
target material) in spite of measuring a residual background a factor
$\simeq$ 3 higher than that of DAMA, thanks to an aggressive
background subtraction.  However such fit was performed using the
energy part of the spectrum for $E_{ee}\ge$ 6 keVee assuming it signal
free, an assumption that is clearly not valid for high--$s$ effective
operators.  Actually, the role of the high--energy part of the
spectrum in momentum--dependent effective models was already partially
understood for $j_\chi$=1/2
in~\cite{xenon100_nreft,bozorgnia_opening_energy}.  In particular
in~\cite{xenon100_nreft} the data from XENON100 were reanalyzed
extending the recoil energy interval up to 240 keVnr.  For this reason
to estimate the present bounds for a xenon target in
Figs.~\ref{fig:markers_plot_mchi_100}
and~\ref{fig:markers_plot_mchi_1000} we included the result of
Ref.~\cite{xenon100_nreft} besides that from
XENON1T~\cite{xenon_2018}~\footnote{The potential importance of the
  large--energy part of the spectrum in DD experiments was also
  discussed in the context of inelastic DM
  models~\cite{inelastic_frontier}.}.

\subsection{Prospects of improvement of present constraints by extending the experimental energy range}
\label{sec:limit_improvements}

In light of what pointed out above, with the exception of PICO--60 the
present bounds on high--multipole effective operators are expected to
be substantially improved by extending the experimental energy windows
beyond the present ones. In order to estimate such improvement, for
each model we have recalculated the bounds on $M/g$ for a xenon,
germanium and sodium iodide target by requiring that the corresponding
expected differential rate in the full energy range where it is
non--vanishing stays below the same residual background levels (in
events/kg/day/keV) achieved by present experiments at lower
energies. Such background levels are shown as horizontal lines in
Fig.~\ref{fig:poster_plot}, and their estimation is explained in
Appendix~\ref{app:exp}.  

\begin{figure}
\begin{center}
  \includegraphics[width=0.7\textwidth]{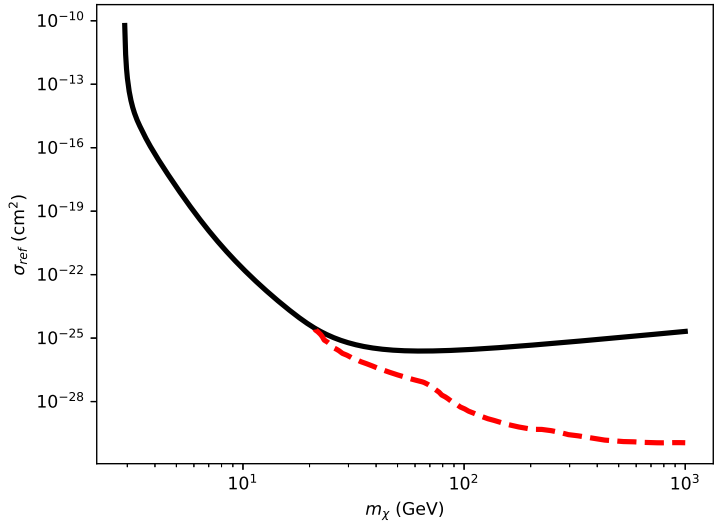}
\end{center}
\caption{Possible improved bounds on the cross section $\sigma_{\rm
    ref} = c_{M,4,4}^2 \mu_{\chi N}^2/\pi$ for the operator
  $\calO_{M,4,4}$ from extending the XENON1T analysis to higher recoil
  energies. Solid (black) line: current XENON1T limit. Dashed (red) line: possible
  improvement.
\label{fig:improved_bounds}}
\end{figure}

The bounds improved in this way are shown in
Figs.~\ref{fig:markers_plot_mchi_100}
and~\ref{fig:markers_plot_mchi_1000} with open markers.  An example of
the improvement in the cross section is presented in
Fig.~\ref{fig:improved_bounds}, where the cross section $\sigma_{\rm
  ref} = c_{M,4,4}^2 \mu_{\chi N}^2/\pi$ for the operator
$\calO_{M,4,4}$ is plotted as a function of the WIMP mass
$m_\chi$. The solid line shows the current limit from XENON1T and the
dashed line shows the possible reach of XENON1T with a high-energy
analysis. The expected improvements on the cross sections are about
one or two orders of magnitude at masses $m_\chi \gtrsim 100$ GeV.

Notice that
experimentally the large energy part of the spectrum is devoid of all
the uncertainties that arise close to threshold, where the efficiency
of applied cuts, the energy scale from light--yields or quenching
factors or the energy resolution are sometimes challenging to
determine. In particular such effects can be safely neglected in the
calculation of the differential rate at high recoil energies.

\section{Conclusions}
\label{sec:conclusions}
While most of the theoretical and experimental work on
detection of particle dark matter has been focused on dark matter
particles that are elementary and have spin 0 or 1/2, there is no
compelling reason for dark matter particles to be elementary, or for
their spin to be limited to 0 and 1/2.

In the present paper we have provided a first systematic and
quantitative discussion of the phenomenology of the non--relativistic
effective Hamiltonian introduced in Ref.~\cite{all_spins} to describe
the nuclear scattering process for a WIMP of arbitrary spin
$j_\chi$.  To this aim we obtained constraints from a representative
sample of present direct detection experiments assuming the
WIMP--nucleus scattering process to be driven by each one of the 44
effective couplings $\calO_{X,s,l}$ that arise for $j_\chi\le$2.

We found that high values of the multipole parameter $s$, related to
powers of the momentum transfer $q$ appearing in the scattering
amplitude, can push the expected differential spectra to recoil
energies $E_R$ much larger than usually assumed, with the largest part
of the signal concentrated at $E_R\gsim$ 100 keV and a peculiar
structure of peaks and minima arising when both the nuclear target and
the WIMP are heavy.  This phenomenology is at strong variance with the
usual WIMP DD paradigm, where the search for new physics is focused on
the lowest--energy part of the spectrum. In particular we have shown
that the present bounds on the effective operators can be
significantly improved by extending the recoil energy intervals to
recoil energies up to $\simeq$ 1 MeV.

A large multipolarity $s$ leads also to a suppression of the expected
rates. In particular we found quantitatively that for $s\le$ 4 the
effective scales probed by direct detection experiments can be suppressed
by up to 5 orders of magnitude compared to the case $s$=0.

It is possible to conceive DM candidates whose interaction with
ordinary matter is driven by the highest multipole moments connected
to the high--rank operators whose phenomenology is the subject of our
analysis. An example is provided by molecules in the dark sector,
where a particularly high symmetry cancels all lower multipoles except
the highest one. In this sense this paper provides the first
phenomenological study of the direct detection of quadrupolar,
octupolar, and hexadecapolar dark matter.

\section*{Acknowledgements}
The research of I.J., S.K. and S.S was supported by the National
Research Foundation of Korea(NRF) funded by the Ministry of Education
through the Center for Quantum Space Time (CQUeST) with grant number
2020R1A6A1A03047877 and by the Ministry of Science and ICT with grant
number 2019R1F1A1052231. GT is supported by a TUM University
Foundation Fellowship. The work of P.G. has been partially supported
by NSF award PHY-1720282 at the University of Utah.

\appendix

\section{Experiments}
\label{app:exp}

In this Appendix we summarize the procedure that we adopted to obtain
the upper bounds discussed in Section~\ref{sec:sensitivities},
categorizing them according to the target: $Xe$, $Ge$, $F$ and
$NaI$. 
\subsection{Xenon target: XENON1T~\&~XENON100}
\label{app:xenon}

For XENON1T we have assumed 7 WIMP candidate events in the range of
3PE $ \le S_1 \le $ 70PE, as shown in Fig.~3 of Ref.~\cite{xenon_2018}
for the primary scintillation signal S1 (directly in Photo Electrons,
PE), with an exposure of 278.8 days and a fiducial volume of 1.3 ton
of xenon (corresponding to a residual background $b_{res,
  XENON1T}\simeq$ 7.7$\times$10$^{-7}$ events/kg/day/keV).  We have
used the efficiency taken from Fig.~1 of~\cite{xenon_2018} and
employed a light collection efficiency $g_1$=0.055; for the light
yield $L_y$ we have extracted the best estimation curve for photon
yields $\langle n_{ph} \rangle /E$ from Fig.~7
in~\cite{xenon_2018_quenching} with an electric field of $90~{\rm
  V/cm}$ (with these assumptions the energy range analyzed by XENON1T
corresponds to 2 keVnr $\lsim E_R\lsim$ 27 keVnr). The energy resolution was
modeled combining a Poisson fluctuation for the observed primary
signal $S_1$ and a Gaussian response of the photomultiplier with
$\sigma_{PMT}=0.5$~\cite{xenon100_resolution}.

Given the relevance of high recoil energies to constrain the effective
models discussed in the present paper, in our analysis we have
included the study of Ref.~\cite{xenon100_high_energy}, where the data
from run II of XENON100 (34 kg × 224.6 live days) were reanalyzed in
the increased recoil energy interval (6.6-240) keV$_{nr}$. To
calculate the bounds we combined the number of observed events $n_k$
and the expected background rates $b_k$ for the 9 bins ($k$=1...9)
listed in Table I of~\cite{xenon100_high_energy} with the
corresponding expected rates $r_k(m_\chi,M/g)$ in the likelihood:

 \begin{equation}
 L(m_\chi,M/g) = 2 \sum_k \left [r_k(m_\chi,M/g) + b_k - n_k ~ log(r_k
   + b_k)\right ]
 \label{eq:lk_xenon100}
 \end{equation}

\noindent and found the upper bounds on $M/g$ imposing the condition
$L(m_\chi,M/g)-L_{min}\le 1.64^2$ at 90\% C.L., with $L_{min}$ the
minimum of $L$. To calculate the expected rates $r_k$ we have directly
convoluted the signal model detector response tables provided for each
of the 9 analysis bins in numerical form
in~\cite{xenon100_high_energy} with the differential rate $dR/dE_R$
calculated in each effective model (see Eq.(B5)
of~\cite{xenon100_high_energy}).

In Section~\ref{sec:limit_improvements} we estimate the improvements
to the present bounds by extending the experimental energy windows
beyond the present ones. To do so for a xenon target we require the
corresponding expected differential rate as given in
Eq.~(\ref{eq:dr_der}) to be below the same residual background $b_{res,
  XENON1T}\simeq$ 7.7$\times$10$^{-7}$ events/kg/day/keV presently
achieved by XENON1T at low energy in the full energy range where it is
non--vanishing. In the calculation of the differential rate at high recoil
energies we assume efficiency equal to unity and neglect the effects
from the energy resolution.

\subsection{Germanium target: SuperCDMS}
The SuperCDMS analysis~\cite{super_cdms_2017} in 2017 observed 1 event
between 4 and 100 keVnr with an exposure of 1690 kg days
(corresponding to a residual background $b_{res, SuperCDMS}\simeq$
6.2$\times$10$^{-6}$ events/kg/day/keV). We have taken the efficiency
from Fig.1 of \cite{super_cdms_2017} and the energy resolution
$\sigma=\sqrt{0.293^2+0.056^2 E_{ee}}$ from \cite{cdms_resolution}. To
analyze the observed spectrum we apply the optimal interval method
\cite{yellin}.

In Section~\ref{sec:limit_improvements} for a germanium target we
follow the same procedure of Section~\ref{app:xenon} using $b_{res,
  SuperCDMS}\simeq$ 6.2$\times$10$^{-6}$ events/kg/day/keV.
\subsection{Fluorine target: PICO--60}

PICO--60 is a threshold experiment utilizing a bubble chamber.  We
analyzed the data obtained with a $C_3F_8$ target~\cite{pico60_2019}
using two thresholds: an exposure of 1404 kg day at threshold
$E_{th}$=2.45 (with 3 observed candidate events and 1 event from the
expected background, implying an upper bound of 6.42 events at
90\%C.L.~\cite{feldman_cousins}) and an exposure of 1167 kg day keV at
threshold $E_{th}$=3.3 keV (with zero observed candidate events and
negligible expected background, implying a 90\% C.L. upper bound of
2.3 events). For the two runs we have assumed the nucleation
probabilities in Fig. 3 of \cite{pico60_2019}.

\subsection{Sodium Iodide target: COSINE--100}

The exclusion plot for COSINE--100~\cite{cosine_nature} relies on a
Montecarlo to subtract the different backgrounds of each of the eight
crystals used in the analysis. In Ref.~\cite{cosine_nature} the amount
of residual background after subtraction is not provided, so we have
assumed a constant background $b$ at low energy (2 keVee$< E_{ee}<$ 8
keVee), and estimated $b$ by tuning it to reproduce the exclusion plot
in Fig.4 of Ref.~\cite{cosine_nature} for the isoscalar
spin-independent elastic case. The result of our procedure yields
$b_{res, COSINE-100}\simeq$0.13 events/kg/day/keVee, which implies a
subtraction of about 95\% of the background.  We take the energy
resolution $\sigma/\mbox{keV}=0.3171
\sqrt{E_{ee}/\mbox{keVee}}+0.008189 E_{ee}/\mbox{keVee}$ averaged over
the COSINE--100 crystals~\cite{cosine_private} and the efficiency for
nuclear recoils from Fig.1 of Ref.~\cite{cosine_nature}. Quenching
factors for sodium and iodine are assumed to be equal to 0.3 and 0.09
respectively, the same values used by DAMA.  In
Section~\ref{sec:limit_improvements} for a sodium iodide target we
follow the same procedure of Section~\ref{app:xenon} using
$b_{res,COSINE-100}$ = 0.13 events/kg/day/keV.

\end{document}